\documentclass[%
	10pt,
	showpacs,%
	amsmath,%
	amssymb,%
	aps,%
	twocolumn,%
	prl%
]{revtex4-1}

\usepackage[english]{babel}
\usepackage[utf8]{inputenc}
\usepackage[T1]{fontenc}
\usepackage{csquotes}

\usepackage[colorlinks=true,citecolor=blue,linkcolor=black,urlcolor=blue]{hyperref}

\usepackage{graphicx}

\usepackage[cspex,bbgreekl]{mathbbol}

\newcommand{\ket}[1]{\left| #1 \right>} % for Dirac bras
\newcommand{\sub}[1]{_{\mathrm{#1}}}
\newcommand{\fig}[1]{Fig.~\ref{#1}}
\newcommand{\eq}[1]{Eq.~\ref{#1}}
\newcommand {\grsim} {\ {\raise-.5ex\hbox{$\buildrel>\over\sim$}}\ }
\newcommand {\lessim} {\ {\raise-.5ex\hbox{$\buildrel<\over\sim$}}\ }

\newcommand{\invisiblesection}[1]{%
  \phantomsection%
  \stepcounter{section}%
  \addcontentsline{toc}{section}{\protect\numberline{\thesection}#1}%
  }

\begin{document}

\title{A Thouless Quantum Pump with Ultracold Bosonic Atoms in an Optical Superlattice}

\author{M.~Lohse$^{1,2}$, C.~Schweizer$^{1,2}$, O.~Zilberberg$^{3}$,  M.~Aidelsburger$^{1,2}$,  I.~Bloch$^{1,2}$}

\affiliation{$^{1}$\,Fakult\"at f\"ur Physik, Ludwig-Maximilians-Universit\"at, Schellingstrasse 4, 80799 M\"unchen, Germany\\
$^{2}$\,Max-Planck-Institut f\"ur Quantenoptik, Hans-Kopfermann-Strasse 1, 85748 Garching, Germany\\
$^{3}$\,Institut für Theoretische Physik, ETH Z\"{u}rich, 8093 Z\"urich, Switzerland\\}

%\date{\today}

%\pacs{XXX}
%TC:endignore
%TC:break abstract

\maketitle
%TC:break default

\invisiblesection{Abstract}
{\bf More than 30 years ago, Thouless introduced the concept of a topological charge pump \cite{Thouless:1983} that would enable the robust transport of charge through an adiabatic cyclic evolution of the underlying Hamiltonian. In contrast to classical transport, the transported charge was shown to be quantized and purely determined by the topology of the pump cycle, making it robust to perturbations \cite{Thouless:1983,Niu:1984}.  On a fundamental level, the quantized charge transport can be connected to a topological invariant, the Chern number, first introduced in the context of the integer quantum Hall effect \cite{Klitzing:1980,Thouless:1982}. A Thouless quantum pump may therefore be regarded as a 'dynamical' version of the integer quantum Hall effect.
Here, we report on the realization of such a topological charge pump using ultracold bosonic atoms that form a Mott insulator in a dynamically controlled optical superlattice potential. By taking in-situ images of the atom cloud, we observe a quantized deflection per pump cycle. We reveal the genuine quantum nature of the pump by showing that, in contrast to ground state particles, a counterintuitive reversed deflection occurs when particles are prepared in the first excited band. Furthermore, we were able to directly demonstrate that the system undergoes a controlled topological phase transition in higher bands when tuning the superlattice parameters.}

\invisiblesection{Topological Charge Pumping}
Charge pumping in solid-state systems has received much attention, mainly due to its potential for realizing novel current standards \cite{Niu:1990,Pekola:2013}, but also for characterizing many-body systems  \cite{Splettstoesser:2005,Marra:2015}. Quantized transport of electrons without an external bias was observed in tunnel junctions with modulated gate voltages \cite{Pothier:1992}, in 1D channels using surface acoustic waves \cite{Talyanskii:1997} and in quantum dots \cite{Blumenthal:2007}. While the latter was used to realize an adiabatic quantum pump \cite{Switkes:1999, Brouwer:1998}, topological charge pumping has so far remained out of reach in condensed matter experiments. In engineered bosonic systems, non-quantized topological pumping of edge states was observed in 1D quasicrystalline photonic waveguide arrays \cite{Kraus:2012, Verbin:2015}. In cold atomic systems, topological Bloch bands have been realized, ranging from the Su-Schrieffer-Heeger~(SSH) model \cite{Atala:2013}, the Hofstadter model in real space \cite{Aidelsburger:2013,Miyake:2013} and synthetic dimensions \cite{Mancini:2015,Stuhl:2015} to the Haldane model \cite{Jotzu:2014}. Their geometric features have been probed using novel interferometric \cite{Atala:2013,Duca:2015} and transport probes \cite{Jotzu:2014,Aidelsburger:2015} that have e.g.\ enabled the direct measurement of the Chern number through bulk topological currents \cite{Aidelsburger:2015}.

Due to their versatility, ultracold atoms in optical superlattices constitute an ideal system for the implementation of quantized topological charge pumps \cite{Romero-Isart:2007, Qian:2011, Wang:2013}. A superlattice is formed by superimposing two lattices with periodicities $d\sub{l}$ and $d\sub{s} = \alpha d\sub{l}$, $\alpha < 1$. This creates a potential $V\sub{s} \sin^2 \left( \pi x / d\sub{s} + \pi/2 \right) + V\sub{l} \sin^2 \left( \pi x / d\sub{l} - \varphi/2 \right) $, where $V\sub{s}$ ($V\sub{l}$) denotes the depth of the short (long) lattice, respectively. The relative position of the lattices is determined by the variable phase $\varphi$. When applying Bloch's theorem, the resulting single-particle Hamiltonian $\hat{\mathcal{H}} (k_x, \varphi)$ is periodic in both the quasi-momentum $k_x \in \left]-\pi/d\sub{l},\pi/d\sub{l}\right]$ and the superlattice phase~$\varphi$.

A cyclic pumping scheme can be realized by adiabatically varying $\varphi$ where one cycle corresponds to a change of $2\pi$, i.e.~moving the long lattice by $d\sub{l}$. A particle which is initially in an eigenstate $\ket{u_n (k_x, \varphi_0)}$ of $\hat{\mathcal{H}}(k_x, \varphi_0)$ in the $n$-th band follows the corresponding instantaneous eigenstate  $\ket{u_{n} (k_x, \varphi)}$. Even with perfect adiabaticity, however, the particle acquires a small admixture of states from other bands, proportional to $\partial_t \varphi$. This gives rise to an anomalous velocity  $\dot{x}_n = \Omega _n \partial_t \varphi$, determined by the Berry curvature  $\Omega_n (k_x, \varphi) = i \left( \langle \partial_{\varphi} u_n | \partial_{k_x} u_n \rangle  - \langle \partial_{k_x} u_n | \partial_{\varphi}  u_n \rangle \right)$ \cite{Karplus:1954, Xiao:2010}. Hence, changing $\varphi$ induces a motion of the particle which, depending on the sign of $\Omega_n$, is either in the same or opposite direction as the moving lattice. The displacement after one cycle is obtained by integrating $\dot{x}_n$ and can in principle take any arbitrary value.

For a filled or homogeneously populated band, however, the average displacement of the entire cloud per cycle can be related to a 2D topological invariant, the Chern number~$\nu_n$ of the pumping process:

\begin{equation}
\label{eq:ChernNumber}
\nu_n = \frac{1}{2 \pi} \int\sub{FBZ} \int_0^{2 \pi} \Omega_n (k_x, \varphi) \ \mathrm{d} \varphi \mathrm{d}k_x
\end{equation}
Here, FBZ denotes the first Brillouin zone of the superlattice. Such a case can be realized with both fermions, by placing the Fermi energy in a band gap, and bosons, by preparing a Mott insulator in the $n$-th band as in our experiment. During one cycle, the cloud's center-of-mass (COM) position changes by $\nu_n d\sub{l}$ and is quantized in units of  $d\sub{l}$ because $\nu_n$ can only take integer values. Unlike in classical systems, this motion can be faster ($\nu_n > 1$) or even opposite ($\nu_n < 0$) compared to the one of the underlying lattice \cite{Wei:2015}. As the displacement is proportional to a topological invariant, it neither depends on the pumping speed, provided adiabaticity still holds, nor on the specific lattice parameters as long as band crossings do not occur. Hence, the transport is highly robust against perturbations such as interaction effects in fermionic systems or disorder \cite{Niu:1984}.

The connection to a Chern topological invariant shows that charge pumping in a 1D superlattice is closely related to the integer quantum Hall effect, where charged particles move in 2D in the presence of a perpendicular magnetic field. Indeed, the adiabatic variation of~$\varphi$ is equivalent to the threading of magnetic flux through a cylinder \cite{Thouless:1983} that generates an electric field in the orthogonal direction and leads to the quantized Hall conductance \cite{Thouless:1982}. A direct analogy between these systems can be made in two limiting cases: the quantum sliding lattice with $V\sub{s} \rightarrow 0$ and the deep tight-binding limit with $V\sub{s} \gg V\sub{l}^2/(4 E\sub{r,s})$. The former matches with the free particle limit of Landau levels that are associated with Chern numbers $\nu_n = +1$. In this limit, one can understand the pumping by following Laughlin's argument \cite{Laughlin:1981}, where the threading of magnetic flux leads to a sliding of the localized oscillator centers. In the latter, one obtains the generalized 1D Harper model \cite{Harper:1955,Roux:2008} which, using dimensional extension \cite{Kraus:2012a,Kraus:2013}, can be mapped onto the 2D Harper-Hofstadter-Hatsugai (HHH) model describing non-interacting particles on a 2D square lattice with a uniform magnetic flux $2 \pi \alpha$ and nearest as well as next-nearest-neighbor hopping \cite{Harper:1955, Azbel:1964, Hofstadter:1976, Hatsugai:1990} (see Supplementary Material). In this mapping, $\varphi$ corresponds to the transverse quasi-momentum~$k_y$. In the following, unless mentioned otherwise, we will focus on the Wannier tunneling regime $V\sub{s} \gtrsim V\sub{l}^2/(4 E\sub{r,s})$ where -- even though the direct mapping to the 1D Harper model breaks down -- the pump is characterized by the same Chern number distribution.

\begin{figure}[t!]
\includegraphics[width=\linewidth]{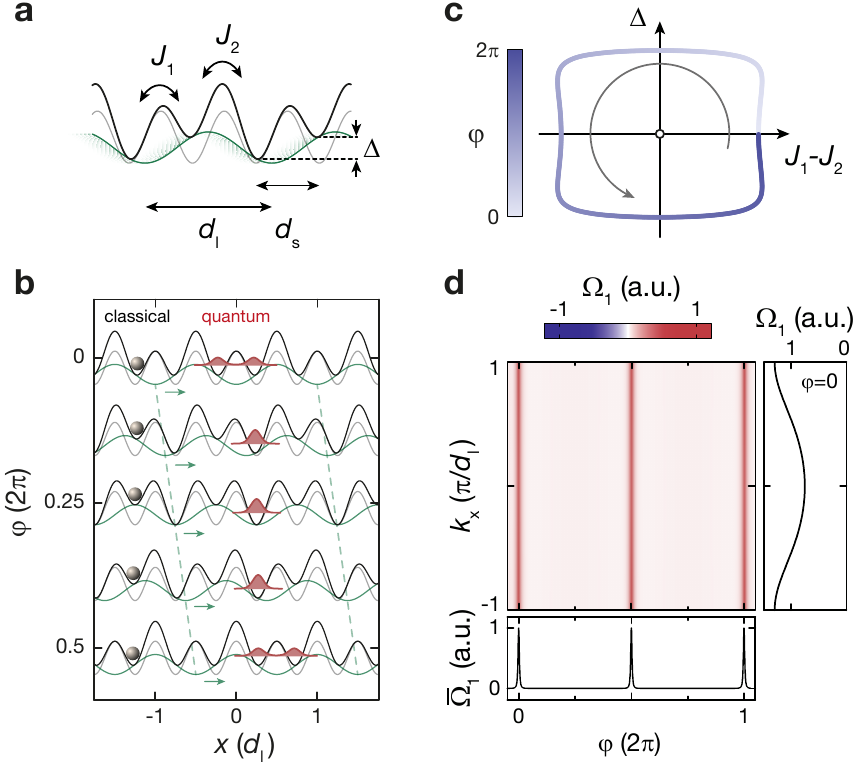}
\vspace{-0.cm} \caption{Topological charge pumping in an optical superlattice. \textbf{(a)} Superlattice potential created by superimposing two lattices with periodicities $d\sub{l}$ and $d\sub{s} = d\sub{l}/2$. Here, $J_1$ and $J_2$ denote the tunnel couplings and $\Delta$ the energy offset between neighboring sites. \textbf{(b)} Evolution of the ground state Wannier function (red) during the first half of a pumping cycle. The long lattice is shifted to the right when increasing the superlattice phase $\varphi$. An atom initially localized in a symmetric superposition on a double well tunnels to the lower lying site and thereby follows the motion of the long lattice in a quantized fashion. A classical particle, on the other hand, stays at a fixed position as the individual sites do not move. \textbf{(c)} Pumping cycle in the  $(J_1-J_2)$-$\Delta$ parameter space. Varying $\varphi$ from 0 to $2\pi$ corresponds to a closed path around the degeneracy point at the origin where $\Delta = 0$ and $J_1 = J_2$. \textbf{(d)} Berry curvature $\Omega_1$ of the lowest band as a function of  $\varphi$  and the quasi-momentum $k_x$ for a lattice with $V\sub{s} = 10\,E\sub{r,s}$ and $V\sub{l} = 20\,E\sub{r,l}$, where $E_{r,i} = h^2/(2m\sub{a} \lambda_i^2)$ denotes the corresponding recoil energy, $m\sub{a}$ the mass of an atom and $\lambda_i$ the respective wavelength. The panel below shows the Berry curvature averaged along $k_x$, as seen by a particle localized in a single double well, which is peaked at the symmetric double well configurations $\varphi = l \pi, l \in \mathbb{Z}$. The graph on the right shows $\Omega_1(k_x)$ for $\varphi = 0$. 
\label{fig:1}}
\end{figure}

\begin{figure*}[t!]
\includegraphics{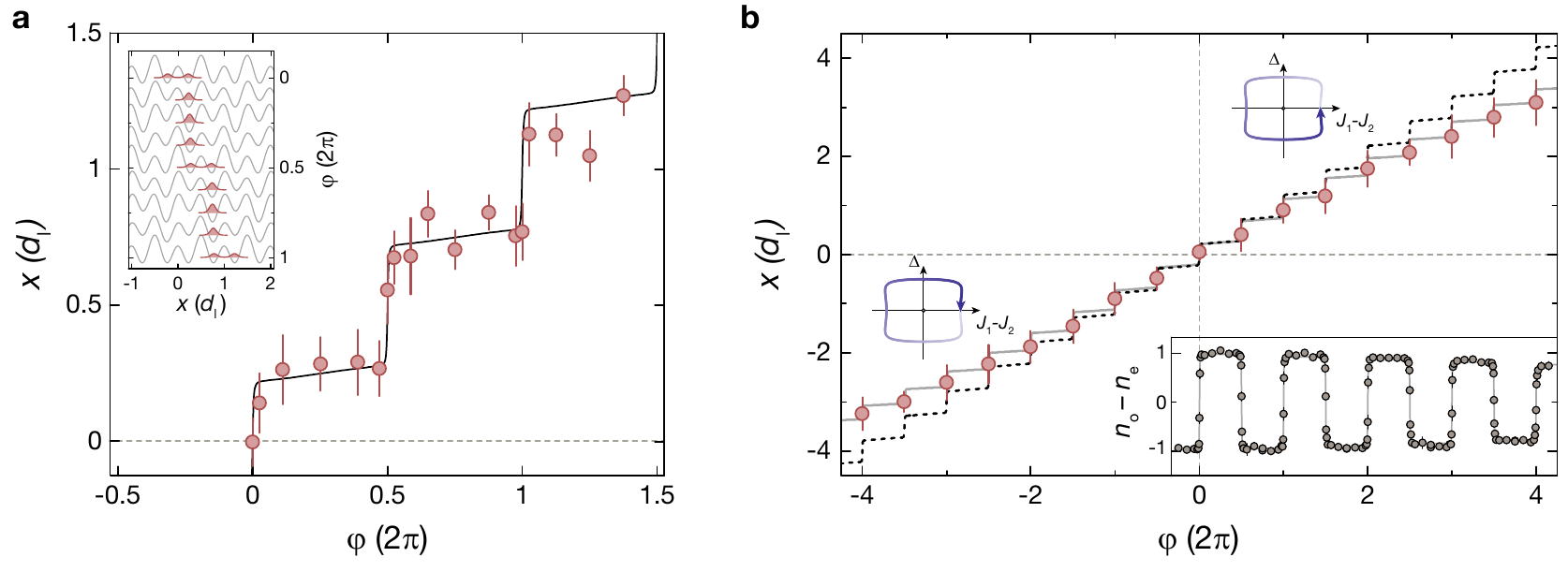}
\vspace{-0.cm}
\caption{Center-of-mass (COM) position of the atom cloud as a function of the pumping parameter $\varphi$ for the lowest band  with $V\sub{s} = 10.0(3)\,E\sub{r,s}$ and $V\sub{l} = 20(1)\,E\sub{r,l}$. \textbf{(a)} Detailed evolution of the measured COM position during a pump cycle. The step-like motion is caused by tunneling of the atoms to the lower lying sites. The solid black line depicts the calculated COM motion of a localized Wannier function.  Each point is the average of ten data sets and the error bar shows the error of the mean. COM positions are determined differentially comparing a sequence with pumping to a reference sequence of the same length, but with constant phase $\varphi = 0$. One data set is obtained by averaging ten images each, taken in alternating order, and subtracting the resulting COM positions. The inset illustrates the motion of a localized Wannier function in the first band during a full pump cycle. \textbf{(b)} COM displacement and site populations for multiple pump cycles in the positive and negative pumping direction. The COM positions are averaged over ten data sets and the error bars depict the standard deviation. The dashed black line shows the ideal motion of a localized Wannier function and the gray line is a fit of the Wannier COM to the data assuming a finite pumping efficiency of 97.9(2)\% per each half of a pump cycle (see Supplementary Material). The inset displays the population imbalance between even and odd sites as a function of $\varphi$ with $n\sub{e}$ ($n\sub{o}$) the fraction of atoms on even (odd) sites. Each data point is the average of five measurements and the error bars indicate the corresponding standard deviation. The gray line is obtained by fitting the calculated even-odd distribution of the Wannier function using the same model as for the COM displacement which yields an efficiency of 98.7(1)\% leading to a slight decrease of the imbalance over time.
\label{fig:2}}
\end{figure*}

In the experiment, we use a superlattice with $d\sub{l} = 2 d\sub{s}$ and two sites per unit cell (\fig{fig:1}a). In the tight-binding limit, it is described by the Rice-Mele Hamiltonian \cite{Rice:1982}

\begin{equation}
\begin{aligned}
\label{eq:RiceMele}
\hat{H}(\varphi) = &- \sum_{m} \left( J_{1}(\varphi) \hat{b}_{m}^{\dagger} \hat{a}^{\phantom \dagger}_{m} + J_{2}(\varphi) \hat{a}_{m+1}^{\dagger} \hat{b}^{\phantom \dagger}_{m} + \mathrm{h.c.} \right) \\
&+ \frac{\Delta(\varphi)}{2} \sum_{m} \left(  \hat{a}_{m}^{\dagger} \hat{a}^{\phantom \dagger}_{m} -  \hat{b}_{m}^{\dagger} \hat{b}^{\phantom \dagger}_{m} \right) 
\end{aligned}
\end{equation}
where $\hat{a}_m^{\dagger}$ ($\hat{a}^{\phantom \dagger}_m$) and $\hat{b}_m^{\dagger}$ ($\hat{b}^{\phantom \dagger}_m$) are the creation (annihilation) operators acting on the even (left) and odd (right) site of the $m$-th unit cell, respectively. $J_1$, $J_2$ denote the tunnel couplings and $\Delta$ the energy offset between neighboring sites. For $\Delta = 0$, this Hamiltonian reduces to the SSH model \cite{Su:1979}.

The mechanism underlying the pumping can also be understood on a microscopic level. Shifting the phase $\varphi$ changes the shape of the potential (\fig{fig:1}b) and modifies $J_1$, $J_2$ and $\Delta$ periodically (\fig{fig:1}c). At $\varphi=0$ ($\Delta = 0$, $J_1 > J_2$), a ground-state particle localized in a single double well is initially in a symmetric superposition of residing on both the left and right sites. With increasing $\varphi$, i.e.\ shifting the long lattice to the right, the double wells are tilted ($\Delta > 0$) and the atom tunnels to the lower lying site on the right. The sign of $J_1 - J_2$ is reversed at $\varphi = \pi/2$, where the tilt is largest, and at $\varphi = \pi$ the lattice forms symmetric double wells again, but shifted by one short lattice constant to the right. The atom, which remained on the lower site for large $\Delta$, delocalizes over the shifted double well  as $\Delta$ becomes comparable to $J_2$ and therefore has moved by $d\sub{l}/2$ during the first half of the pumping cycle. In the second half, the same procedure is repeated, but shifted by one site. After one cycle, the lattice configuration is identical to the starting point, but the atom ends up in the double well next to the initial one. In contrast to this, a classical particle would not move because the positions of the individual sites do not change. This illustrates the importance of quantum tunneling for the pumping.

During the pump cycle, the system moves along a closed trajectory in the ($J_1-J_2)$-$\Delta$ parameter space. It encircles the degeneracy point at $\Delta = 0 $ and $J_1 = J_2$, where the two bands of the Rice-Mele model touch, and smoothly connects the topologically distinct phases $J_1 <  J_2$ and $J_1 > J_2$ of the SSH model  (\fig{fig:1}c). The Berry curvature and thus the motion of the atoms is peaked around $\varphi = l \pi, l \in \mathbb{Z}$ where the tilt changes sign and the atoms tunnel to the neighboring sites (\fig{fig:1}d). The Chern numbers of the pump cycles in the two bands of this model are $\nu_1 = +1$ and $\nu_2 = -1$, giving the same Chern number distribution as the HHH model with $\alpha = 1/2$ \cite{Hatsugai:1990}. Their sum vanishes because these bands emerge from the topologically trivial lowest Bloch band of the short lattice.

The experimental sequence starts by preparing an $n=1/2$ Mott insulator of $^{87}$Rb atoms in a 3D optical lattice with at most one atom per unit cell in the ground state of symmetric double wells with $\varphi = 0$ and $J_1 \gg J_2$ (see Methods).  Due to the large on-site interaction, each atom is localized on an individual double well, resulting in a homogeneous delocalization over the entire first Brillouin zone. The pumping is performed by adiabatically shifting the phase $\varphi$ of the long lattice (see Methods) and the resulting motion of the atoms is tracked by measuring the COM position of the cloud in situ. The displacement during one cycle is indeed quantized and occurs in steps (\fig{fig:2}a) -- unlike the underlying linear motion of the long lattice. The cloud moves by one lattice constant $d\sub{l}$ per cycle as expected for $\nu_1 = +1$ and the steps appear around $\varphi = l \pi, l \in \mathbb{Z}$, where the atoms tunnel from one side of the double wells to the other. When performing multiple cycles, the cloud keeps moving to the right whereas it propagates in the opposite direction for the reversed pumping direction $\varphi < 0$ (\fig{fig:2}b). The small deviation from the expected displacement for the motion of ideal Wannier functions can be attributed to a finite pumping efficiency due to non-adiabatic band transitions and the additional trapping potential.

\begin{figure}[t!]
\includegraphics{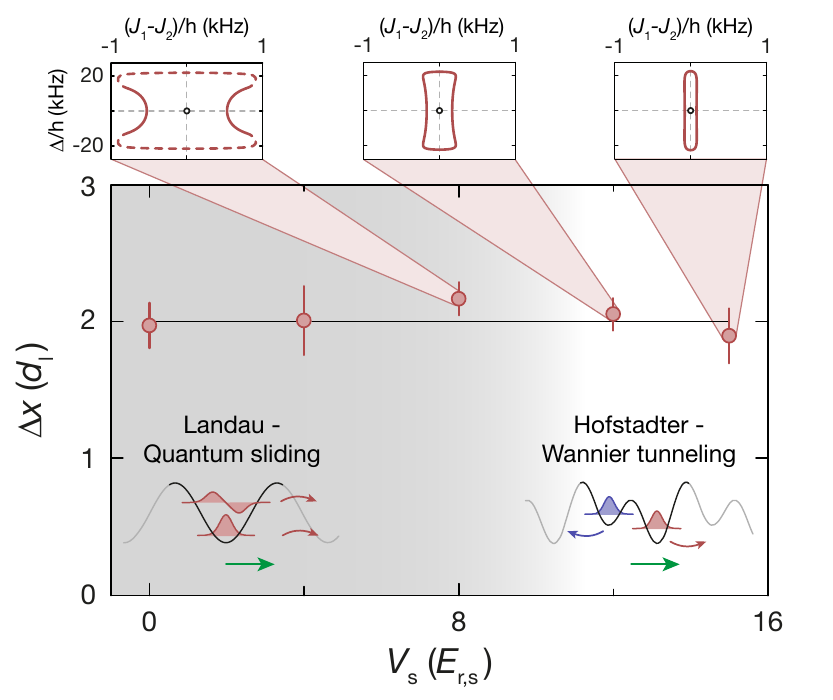}
\vspace{-0.cm}
\caption{Transition from a quantum sliding lattice to the Wannier tunneling limit for the lowest band. Differential deflection $\Delta x$ after one pump cycle for positive and negative pumping direction for various lattice depths $V\sub{s}$ at $V\sub{l} = 25(1)\,E\sub{r,l}$. Each point consists of ten data sets comparing the COM position of ten averaged images for both directions. The error bars depict the error of the mean. For the data points in the tight-binding regime, the insets show the corresponding pump cycles in the ($J_1-J_2)$-$\Delta$ parameter space. For $V\sub{s} = 8\,E\sub{r,s}$, the two-band model breaks down for large tilts such that $J_1$, $J_2$ and $\Delta$ are not well-defined. The dashed line therefore connects the points where the gap between the second and third band becomes smaller than $10J_1$ for $\varphi = 0$.  
\label{fig:3}}
\end{figure}

The step-like transport behavior can also be observed in site-resolved band mapping measurements (inset of \fig{fig:2}b) which determine the number of atoms on even and odd sites. As for the COM position, a step occurs in the even-odd distribution whenever a symmetric double well configuration is crossed at $\varphi = l \pi, l \in \mathbb{Z}$. Using the measured even-odd fractions, one can estimate the transfer efficiency, i.e.\ the fraction of atoms transferred from site $i$ to $i+1$ at each step. This is equivalent to the fraction staying in the lowest band during one half of the pumping cycle and allows to quantify the adiabaticity of the pumping protocol. From our data we obtain an efficiency of 98.7(1)\% (see Supplementary Material). With the same model, the ideal COM displacement can be fitted to the measured positions which yields an efficiency of $97.9(2)\%$ with the small additional reduction most likely being caused by the trap.

Due to the topological nature of the pumping, the displacement per cycle for the lowest band does not depend on the path in the $(J_1 - J_2)$-$\Delta$ plane as long as it encompasses the degeneracy point. Moreover, it is independent of $V\sub{s}$ since the sliding lattice and the tight-binding Thouless pump are topologically equivalent for the first band and connected by a smooth crossover without closing the gap to the second band. To verify this, we measured the deflection of the cloud with $V\sub{l} = 25(1)\,E\sub{r,l}$ for various values of $V\sub{s}$. For all parameters, the resulting displacements are consistent within the error bars (\fig{fig:3}).

\begin{figure}[t!]
\includegraphics{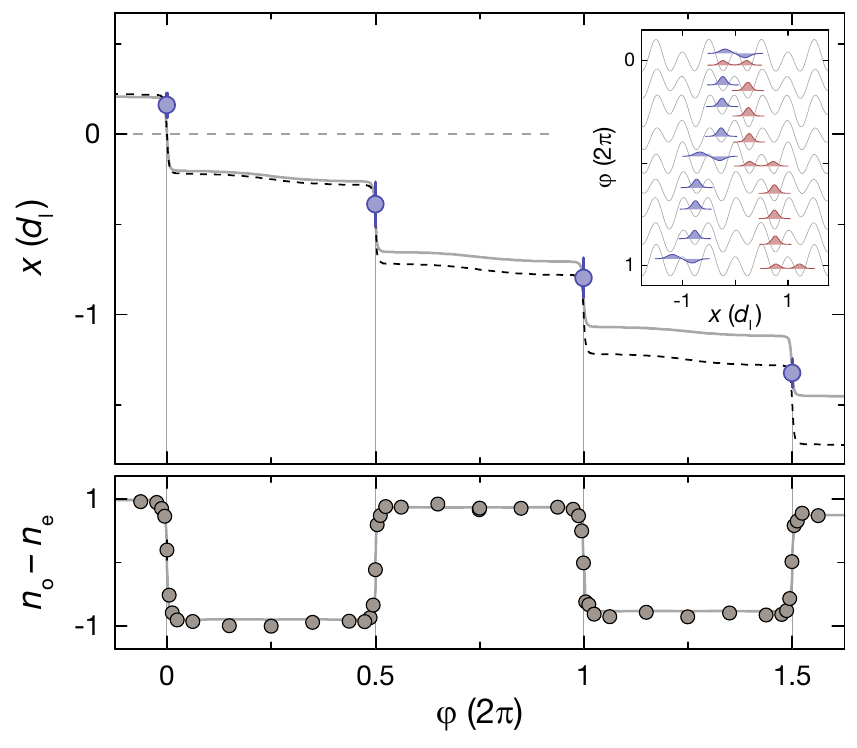}
\vspace{-0.cm}
\caption{Cloud displacement and site occupations for the first excited band with $V\sub{s} = 10.0(3)\,E\sub{r,s}$ and $V\sub{l} = 20(1)\,E\sub{r,l}$. The main plot shows the evolution of the COM position for up to 1.5 pump cycles. The data points are averaged over ten data sets with the errors being the error of the mean. The dashed black line indicates the motion of a localized Wannier function of the second band and the gray line is a fit to the data using the same model as in \fig{fig:2}b, giving a transfer efficiency of 97(2)\%. The inset illustrates the evolution of the Wannier functions of the first and second band during a pump cycle, showing a deflection in opposite directions. The lower plot shows the imbalance of the fraction of atoms on even ($n\sub{e}$) and odd sites ($n\sub{o}$) averaged over 3-6 measurements each with the error bars depicting the standard deviation. The gray line is a fit of the even-odd distribution of the corresponding Wannier functions with a pumping efficiency of 96.7(3)\% (see Supplementary Material).
\label{fig:4}}
\end{figure}

The excited band in the Rice-Mele model exhibits counter-propagating charge pumping with $\nu_2 = -1$, i.e.\ the atoms are expected to move in the opposite direction as the long lattice. This underlines the pump's intrinsic quantum mechanical character as such a motion could not occur for classical particles. To study this, the atoms were prepared in the second band at $\varphi = 0$ (see Methods) and pumping was performed with identical parameters as for the first band in \fig{fig:2}.\ When moving the lattice to the right ($\varphi > 0$), the cloud indeed shifts to the left with $x < 0$ (\fig{fig:4}). This is further confirmed by the measured site occupations, showing that the behavior is exactly reversed compared to the lowest band. While the first band is localized on the lower site for large $\Delta$, atoms in the second band are found on the upper site and are therefore transported in the opposite direction. The slightly larger deviation from theory is mostly due to the finite lifetime of atoms in the higher band. This leads to a lower transfer efficiency of 96.7(3)\% obtained from the site populations, in agreement with the value from the fit to the COM positions of 97(2)\%  (see Supplementary Material).

\begin{figure}[t!]
\includegraphics{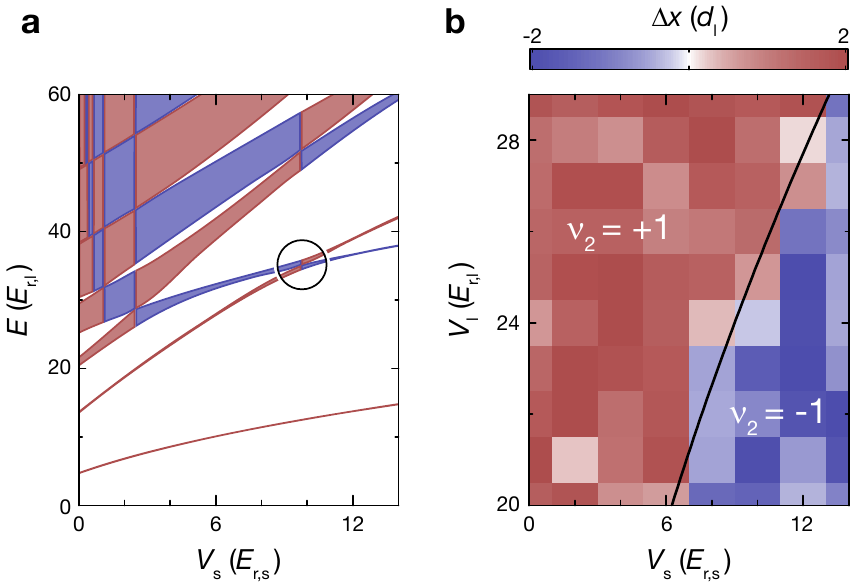}
\vspace{-0.cm}
\caption{Topological phase transition in the first excited band. \textbf{(a)} Band structure at $\varphi = 0.5\pi$, where the band crossings occur, and Chern number distribution versus $V\sub{s}$ with $V\sub{l} = 25\,E\sub{r,l}$. The shaded areas illustrate the width of the Bloch bands, which were calculated by numerical diagonalization of $\hat{\mathcal{H}}(k_x, \varphi)$. The color indicates the Chern number of the corresponding pump cycle with red being $+1$ and blue $-1$. Starting from alternating Chern numbers in the Wannier tunneling limit for large $V\sub{s}$, the negative Chern numbers successively propagate towards higher bands in a series of topological phase transitions when lowering $V\sub{s}$, giving rise to a uniform distribution with $\nu_n = +1$ in the Landau limit of a sliding long lattice. The circle highlights the crossing of the 2nd and 3rd band leading to the topological transition studied in b. \textbf{(b)} Differential deflection $\Delta x$ between single pump cycles in opposite directions for the second band as a function of the lattice depths $V\sub{s}$ and $V\sub{l}$. At small $V\sub{s}$, $\nu_2 = +1$ and atoms move in the same direction as the lattice, but as $V\sub{s}$ increases a topological transition occurs at $V\sub{s} =V\sub{l}^2/(4 E\sub{r,s})$, indicated by the black line, where $\nu_2$ suddenly changes to $-1$ and the direction of motion is reversed. Each square corresponds to one data set averaged over 10-12 pairs of images.
\label{fig:5}}
\end{figure}

By varying $V\sub{s}/V\sub{l}^2$, one can study the transition between the topologically very different regimes of the sliding long lattice with $V\sub{s} = 0\,E\sub{r,s}$ and the Wannier tunneling limit for $V\sub{s} > V\sub{l}^2/(4 E\sub{r,s})$. For the latter, the Chern numbers of the bands alternate between $\nu_{2n+1} = +1$ and $\nu_{2n} = -1$ for $d\sub{l} = 2 d\sub{s}$, like in the HHH model with a flux of $\pi$. This causes the opposite deflections for the first and second band observed in \fig{fig:2} and \fig{fig:4}. In the other limit $V\sub{s} \rightarrow 0\,E\sub{r,s}$, however, each band corresponds to a single Landau level with $\nu_n = +1$. Between these two limiting cases, as $V\sub{s}$ is decreased, an infinite series of topological phase transitions occurs where two bands touch and exchange their Chern numbers (\fig{fig:5}a). Thereby the negative $\nu_n$ are successively transferred to higher bands and they become positive for the lower bands.

While the deflection in the first band is independent of $V\sub{s}$, there is a transition for the second band where the gap to the third band closes and $\nu_2$ changes from $-1$ to $+1$. This transition can be mapped out by measuring the displacement of the cloud as a function of the lattice depths $V\sub{s}$ and $V\sub{l}$ (\fig{fig:5}b). The direction of the motion reverses when crossing the transition, which occurs at $V\sub{s} =V\sub{l}^2/(4 E\sub{r,s})$ in the tight-binding limit. For small $V\sub{s}$, atoms in the second band move in the same direction as the lattice, whereas they move in the opposite directionin the Wannier tunneling regime.

In conclusion, we have demonstrated the implementation of a topological charge pump using ultracold atoms. Unlike previous experiments studying non-quantized pumping of edge states, we observe a quantized response in the bulk. Combining such a pump with novel techniques for engineering optical potentials at the single-site level \cite{Weitenberg:2011, Preiss:2015} would allow for a direct observation of edge states in finite systems \cite{Kitagawa:2012} and to study their transport properties \cite{Kraus:2012}. Furthermore, by adjusting the ratio of the lattice constants $\alpha = d\sub{s}/d\sub{l}$, one can realize a wide variety of commensurate and incommensurate superlattices where the Chern number of the lowest band can in principle take arbitrary integer values. In addition to the negative deflection for $\nu_n < 0$ shown here, another counterintuitive case of charge pumping can occur in these systems where the atoms move faster than the long lattice for $\nu_n > 1$ \cite{Wei:2015}.  By adding a spin degree of freedom, the pumping scheme can be used to implement the $Z_2$ spin pump \cite{Shindou:2005, Fu:2006} in a spin-dependent superlattice \cite{Lee:2007}. Moreover, extending the Thouless pump to 2D systems would enable the realization of an analogon of the 4D integer quantum Hall effect \cite{Zhang:2001, Kraus:2013}. \\
\invisiblesection{Notes and Acknowledgments}
\newline
\paragraph{Note} Recently, we became aware of similar work by S. Nakajima et al. \cite{Nakajima:2015} implementing the Thouless pump with fermionic atoms. \\
\newline

We acknowledge insightful discussions with F. Grusdt and S. Kohler. This work was supported by NIM and the EU (UQUAM, SIQS). M. L. was additionally supported by ExQM and O. Z. by the Swiss National Science Foundation.

\section*{Methods}

\small

{\bf Initial state preparation in the first band}
All sequences started by loading an $n=1$ Mott insulator of about 3000 $^{87}$Rb atoms in the lowest band of a 3D optical lattice. The lattice was created by three orthogonal standing waves with wavelengths $\lambda\sub{l} = 1534$\,nm for the long lattice along the $x$-direction and $\lambda_{y} = 767$\,nm and $\lambda_{z} = 844$\,nm along the $y$- and $z$-axes, respectively. They were ramped up in 150\,ms to a depth of $V_i = 30(1)\,E_{r,i}, i \in \{l,y, z \}$. The superlattice potential was created by adding another lattice with $\lambda\sub{s} = \lambda\sub{l}/2$ along the $x$-direction. To prepare the atoms in the lowest band of the superlattice, the lattice sites along the $x$-axis were split symmetrically with $\varphi = 0.00(1)\pi$ by ramping up the short lattice to $V\sub{s} = 10.0(3)\,E\sub{r,s}$ within 10\,ms while simultaneously lowering $V\sub{l}$ to $20(1)\,E\sub{r,l}$ such that $J_1 \gg J_2$. It was verified with direct band mapping measurements that the atoms are homogeneously distributed over the entire band. 

{\bf Initial state preparation in the second band}
For the preparation in the second band, the splitting was instead performed at $\varphi = 0.11(1)\pi$ and $V\sub{s}$ was ramped to 30(1)\,$E\sub{r,s}$. The phase was then changed non-adiabatically to $\varphi = -0.08(1)\pi$ in 20\,ms to transfer all atoms to the excited band. After that, the lattices along $x$ were lowered to $V\sub{s} = 10.0(3)\,E\sub{r,s}$ and $V\sub{l}$ to $20(1)\,E\sub{r,l}$, respectively, in 2\,ms and the phase was moved to $\varphi = 0.00(1) \pi$ in 10\,ms. This brings the atoms into the excited state of symmetric double wells with almost perfect efficiency.

{\bf Sequence for pumping}
The pumping cycle was implemented experimentally by slightly changing the laser frequency and thereby shifting the phase $\varphi$ of the long lattice along the $x$-direction. For this, two separate lasers were used with one laser covering the range from $-0.50(1)\pi$ to $0.62(1)\pi$ and the second laser from $0.62(1)\pi$ to $1.50(1)\pi$ (see Supplementary Material). 

\normalsize

%%%%%%%% Bibliography %%%%%%%%
%\bibliographystyle{naturemag}
\bibliographystyle{bosons}

%%%%%%%% Supplementary Material %%%%%%%%
\onecolumngrid
\clearpage
\begin{center}
\noindent\textbf{Supplementary Material for:}
\\\bigskip
\noindent\textbf{\large{A Thouless Quantum Pump with Ultracold Bosonic Atoms in an Optical Superlattice}}
\\\bigskip
M.~Lohse$^{1,2}$, C.~Schweizer$^{1,2}$, O.~Zilberberg$^{3}$,  M.~Aidelsburger$^{1,2}$,  I.~Bloch$^{1,2}$
\\\vspace{0.1cm}
\small{$^1$ \emph{Fakult\"at f\"ur Physik, Ludwig-Maximilians-Universit\"at,\\ Schellingstrasse 4, 80799 M\"unchen, Germany}}\\
\small{$^2$ \emph{Max-Planck-Institut f\"ur Quantenoptik,\\ Hans-Kopfermann-Strasse 1, 85748 Garching, Germany}}\\
\small{$^3$ \emph{Institut für Theoretische Physik, ETH Z\"{u}rich, 8093 Z\"urich, Switzerland}}
\end{center}
\bigskip
\bigskip
\twocolumngrid

%%%%%%%% Begin Body %%%%%%%%

\renewcommand{\thefigure}{S\the\numexpr\arabic{figure}-10\relax}
 \setcounter{figure}{10}
\renewcommand{\theequation}{S.\the\numexpr\arabic{equation}-10\relax}
 \setcounter{equation}{10}
 \renewcommand{\thesection}{S.\Roman{section}}
\setcounter{section}{10}
\renewcommand{\bibnumfmt}[1]{[S#1]}
\renewcommand{\citenumfont}[1]{S#1}

%%%%%%%% Mapping to the Hofstadter model in the tight-binding regime %%%%%%%%
\section{Mapping to the Hofstadter model in the deep tight-binding regime}

In the tight-binding approximation, the dynamics of ultracold atoms in an optical superlattice potential with arbitrary lattice constants $d\sub{l}$ and $d\sub{s} = \alpha d\sub{l}$, $\alpha < 1$ can be described by the following Hamiltonian: 

\begin{equation}
\begin{aligned}
\label{eq:SLHamiltonian}
\hat{H}(\varphi) = &- \sum_{m} \left( \left(J_0 + \delta J^{\phantom \dagger}_m (\varphi) \right) \hat{a}_{m+1}^{\dagger} \hat{a}^{\phantom \dagger}_{m} + \mathrm{h.c.} \right) \\
&+   \sum_{m}  \Delta^{\phantom \dagger}_m(\varphi) \hat{a}_{m}^{\dagger} \hat{a}^{\phantom \dagger}_{m}
\end{aligned}
\end{equation}

\noindent Here,  $\hat{a}_m^{\dagger}$ ($\hat{a}^{\phantom \dagger}_m$) is the creation (annihilation) operator acting on the $m$-th lattice site and $J$ denotes the hopping amplitude between neighboring sites in the short lattice. $\delta J_m$ is the modulation of the hopping amplitude and $\Delta_m$ the on-site energy of the $m$-th site, both of which are induced by the long lattice and depend on $\varphi$. For  $V\sub{s} \gg V\sub{l}^2/(4 E\sub{r,s})$, \eq{eq:SLHamiltonian} reduces to a generalized 1D Harper model \cite{S:Harper:1955,S:Roux:2008}  with 

\begin{equation}
\begin{aligned}
\label{eq:HarperParameters}
\delta J_m (\varphi) &= \frac{\delta J}{2}         \cos \bigl( 2\pi \alpha m       - \varphi \bigr) \\
\Delta_m (\varphi)   &= -\frac{\Delta}{2} \cos \bigl( 2\pi \alpha (m-1/2) - \varphi \bigr)
\end{aligned}
\end{equation}

\begin{figure}[thb]
\begin{center}
\includegraphics[scale=1]{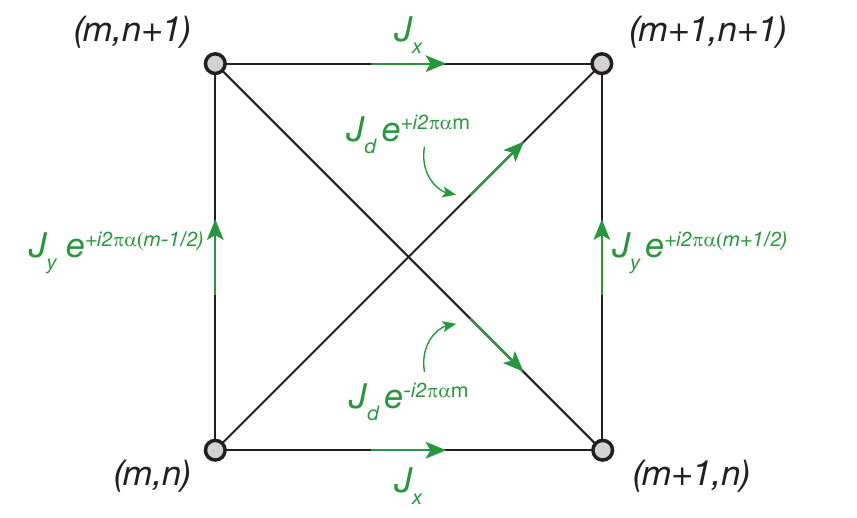}
\caption{Square lattice in the presence of a uniform magnetic field with nearest-neighbor hopping $J_x = J_0, J_y = \Delta/4$ and next-nearest-neighbor hopping $J_d = \delta J /4$ using the gauge of \eq{eq:3hHamiltonian}.}
\label{fig:S1}
\end{center}
\end{figure}

Following the approach of dimensional extension \cite{S:Kraus:2012a}, the periodic parameter $\varphi$ can be regarded as an extra dimension, i.e. a realization for a given $\varphi$ corresponds to a 1D subset of a 2D system. The pumping then samples the full 2D parameter space which is the reason why the displacement can be related to a 2D topological invariant. The 2D Hamiltonian onto which the 1D superlattice maps in this limit can be found by thinking of the 1D Hamiltonian as a single Fourier component of the 2D one and performing an inverse Fourier transform. For this, we relabel the creation and annihilation operators as $\hat{a}^{\dagger}_{m, \varphi}$ and $\hat{a}^{\phantom \dagger}_{m, \varphi}$ and define a 2D Hamiltonian

\begin{equation}
\label{eq:2DHamiltonian}
\hat{H}\sub{2D} = \frac{1}{2\pi} \int_0^{2\pi} \hat{H}(\varphi) \,\mathrm{d\varphi} 
\end{equation}

\noindent Expanding $\hat{a}^{\dagger}_{m, \varphi}$ and $\hat{a}^{\phantom \dagger}_{m, \varphi}$ into their Fourier components, $\hat{a}^{\dagger}_{m,\varphi} = \sum_n e^{i \varphi n} \hat{a}^{\dagger}_{m,n}$ and $\hat{a}^{\phantom \dagger}_{m,\varphi} = \sum_n e^{-i \varphi n} \hat{a}^{\phantom \dagger}_{m,n}$, gives 

\begin{equation}
\begin{aligned}
\label{eq:3hHamiltonian}
\hat{H}\sub{2D} = - &\sum_{m,n}  J_0  \hat{a}_{m+1,n}^{\dagger} \hat{a}^{\phantom \dagger}_{m,n} \\
- &\sum_{m,n} \frac{\Delta}{4}   e^{+i 2\pi \alpha (m - 1/2)}  \hat{a}_{m,n+1}^{\dagger} \hat{a}^{\phantom \dagger}_{m,n}  \\
- &\sum_{m,n} \frac{\delta J}{4} e^{+i 2\pi \alpha m}  \hat{a}_{m+1,n+1}^{\dagger} \hat{a}^{\phantom \dagger}_{m,n} \\
- &\sum_{m,n} \frac{\delta J}{4} e^{-i 2\pi \alpha m}  \hat{a}_{m+1,n}^{\dagger} \hat{a}^{\phantom \dagger}_{m,n+1} \\
+ &\mathrm{\,h.c.}
\end{aligned}
\end{equation}

\noindent which is precisely the 2D Harper-Hofstadter-Hatsugai model describing non-interacting particles on a 2D square lattice in the presence of a uniform magnetic flux $2 \pi \alpha$ per square plaquette with nearest and next-nearest neighbor tunneling as illustrated in \fig{fig:S1} \cite{S:Hatsugai:1990}. Under this mapping, the superlattice phase $\varphi$ corresponds to the transverse quasi-momentum $k_y$ and in this sense the pumping protocol is equivalent to performing Bloch oscillations in the 2D lattice by applying a gradient. For our particular case of $d\sub{s} = d\sub{l}/2$, one obtains a flux of $\pi$ per plaquette.

%%%%%%%% Mapping of the sliding lattice to Landau levels %%%%%%%%
\section{Mapping of the sliding lattice to Landau levels}

In the opposite limit of the sliding lattice with $V\sub{s} \rightarrow 0$, a similar analogy can be made to the Landau levels of a free particle in an external magnetic field in 2D. The Hamiltonian of a single particle in the long lattice reads

\begin{equation}
\label{eq:LongLatticeHamiltonian}
\hat{H}\sub{l} = \frac{\hat{p}_x^2}{2 m\sub{a}} + V\sub{l} \sin^2 \left( \pi \alpha \hat{x} / d\sub{s} - \varphi/ 2 \right)
\end{equation}

\noindent For a deep lattice and when only considering the lowest bands, the lattice potential can be expanded in orders of $\hat{x}- x_m$ around each lattice site $x_m$ (i.e. the minimum of the potential). In second order this gives: 

\begin{equation}
\label{eq:LLHamiltonianExp}
\hat{H}\sub{l} = \frac{\hat{p}_x^2}{2 m\sub{a}} + V\sub{l} \frac{\pi^2 \alpha^2}{d\sub{s}^2} \sum_m (\hat{x} - x_m)^2
\end{equation}

\noindent Neglecting the coupling between neighboring sites, this splits into a series of decoupled Hamiltonians for each lattice site, each of which is a 1D harmonic oscillator. 

On the other hand, the motion of a particle with charge $q$ in an external magnetic field $\mathbf{B} = B \bf{e}_z$ along the $z$-direction is described by the Hamiltonian 

\begin{equation}
\begin{aligned}
\label{eq:LandauHamiltonian}
\hat{H}\sub{LL} &= \frac{1}{2m\sub{a}} \left( \mathbf{\hat{p}} - q \hat{\mathbf{A}} \right)^2 \\
								&= \frac{\hat{p}\sub{x}^2}{2m\sub{a}} +  \frac{1}{2m\sub{a}} \left( \hat{p}\sub{y} - q B\hat{x} \right)^2
\end{aligned}
\end{equation}

\noindent with the vector potential $\hat{\mathbf{A}} = B \hat{x} \bf{e}\sub{y}$ given in the Landau gauge. As the above Hamiltonian commutes with the operator for the transverse momentum $\hat{p}_y$, one can define a common set of eigenstates of both $\hat{H}\sub{LL}$ and $\hat{p}_y$. For a given state with momentum $\hbar k_y$, \eq{eq:LandauHamiltonian} can also be rewritten as a 1D harmonic oscillator

\begin{equation}
\label{eq:ReducedLandauHamiltonian}
\hat{H}\sub{LL} = \frac{\hat{p}\sub{x}^2}{2m\sub{a}} +  \frac{1}{2} m\sub{a} \omega^2\sub{c} \left( \hat{x} - \frac{\hbar k\sub{y}}{m\sub{a} \omega\sub{c}}\right)^2
\end{equation}

\noindent where $\omega\sub{c} = qB/m\sub{a}$ is the cyclotron frequency. Comparing this with \eq{eq:LLHamiltonianExp}, one can see that the state in the $n$-th band localized at the lattice site $x_m$ corresponds to the state in the $n$-th Landau level with a transverse momentum $\hbar k_y = m\sub{a} \omega\sub{c} x_m$. As in the tight-binding regime, the number of magnetic flux quanta per unit cell of the (non-existent) short lattice is given by the ratio of the lattice constants $\alpha = d\sub{s}/d\sub{l}$ in this mapping.

%%%%%%%% Topological phase transition %%%%%%%%%
\section{Topological phase transition}
For $d\sub{s} = d\sub{l}/2$, the phase transition to the Wannier tunneling regime, where the pump is characterized by the same Chern number distribution as the 2D HHH model, occurs when the second and third band cross at the staggered configuration $\varphi = \pi/2$ during the pumping cycle. In the tight binding limit, the crossing point can be determined by comparing the energy of the first excited state on the lower site with the one of the ground state on the higher site. For a sufficiently deep short lattice, they can be approximated by harmonic oscillator states with $E_{n} = (n+1/2) \hbar \omega$. The on-site trapping frequency~$\omega$ is given by
\begin{equation}
\omega=\sqrt{\frac{2 \pi^{2}V\sub{s}}{m\sub{a} d\sub{s}^2}}=\frac{2}{\hbar}\sqrt{V\sub{s}E\sub{r,s}}
\label{eq:trappingfrequency}
\end{equation}
At $\varphi = \pi/2$, the tilt is largest with~$\Delta=V\sub{l}$ and the bands touch if $\Delta=\hbar\omega$. Hence, the transition occurs at
\begin{equation}
V\sub{l}=2\sqrt{V\sub{s}E\sub{r,s}}
\label{eq:crossing1st2ndband}
\end{equation}

\noindent A similar derivation can be made for $d\sub{s} = d\sub{l}/(2j), j \in \mathbb{N}$ which gives the same transition point. For all other ratios, \eq{eq:crossing1st2ndband} gives a lower bound of $V\sub{l} (V\sub{s})$ and the transition will in general take place at slightly larger $V\sub{l}$.

%%%%%%%% Experimental sequence %%%%%%%%%
\section{Experimental sequence}
\begin{figure}
\includegraphics[width=\linewidth]{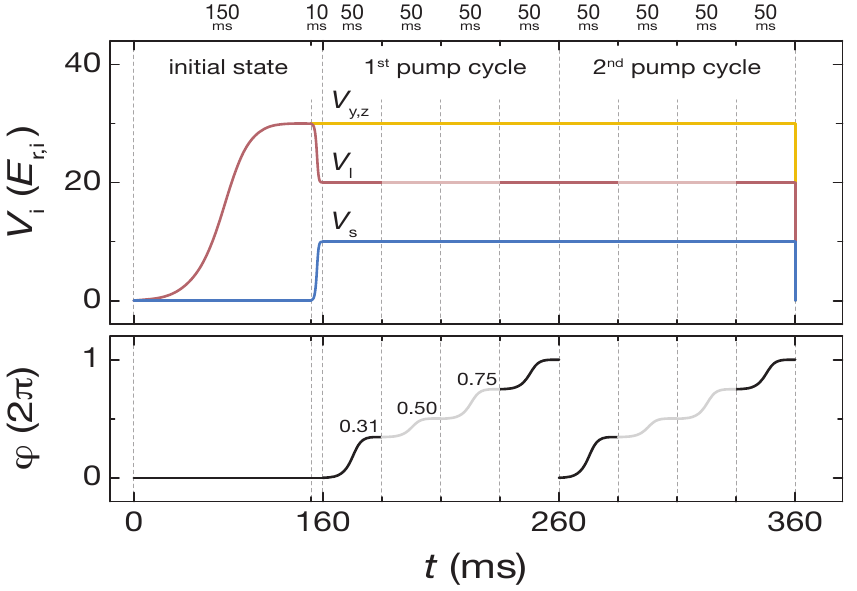}
\vspace{-0.5cm} 
\caption{Schematic of the experimental sequence with the lattice and phase ramps for the initial state preparation as well as for two pumping cycles. The light and dark lines indicate the two different lasers used sequentially to overcome tuning range limitations. The phase~$\varphi$ is shown modulo $2\pi$.}
\label{fig:S2}
\end{figure}

Figure~\ref{fig:S2} illustrates the detailed experimental sequence for the initial state preparation of the atoms in the lowest band of the superlattice potential and two exemplary pump cycles. As discussed in the Methods of the main text, the sequence started by loading an $n=1$ Mott insulator in $150$\,ms in a 3D optical lattice created by the long lattice and two orthogonal standing waves with depth $V_i = 30(1)\,E_{r,i}, i \in \{l,y,z\}$. Subsequently, the long lattice sites were split symmetrically by ramping up the overlapped short lattice with $\lambda\sub{s}=\lambda\sub{l}/2$ to $V\sub{s} = 10.0(3)\,E\sub{r,s}$ at $\varphi = 0$. Simultaneously, the long lattice was lowered to $V\sub{l}=20(1)\,E\sub{r,l}$ such that $J_1 \gg J_2$.

For the pumping, the phase of the long lattice was moved by changing the laser frequency slightly. Due to the limited tuning range of a single laser, a successive hand over between two independent laser systems (indicated with light and dark lines in Fig.~\ref{fig:S2} for the different lasers) was implemented in order to extend the pumping range to arbitrary values. Each pump cycle consists of four segments with $s$-shaped phase ramps  $\varphi \in [0, 0.62\pi],[0.62\pi, \pi],[\pi, 1.5\pi]$  and $[1.5\pi, 2\pi]$ of 50\,ms duration to minimize the probability for non-adiabatic transitions to higher bands at the symmetric configurations $\varphi = l \pi, l \in \mathbb{Z}$. For the measurements in Fig. 3 and  Fig.~5 of the main text, the ramp time was scaled with $V\sub{s}$ as the tunneling rates and thus the band gap decrease with larger $V\sub{s}$. The switching between the lasers at $0.62\pi$ and $1.5\pi$ was done instantaneously within a 3\,ms hold time between the ramps. It was confirmed experimentally that this switch does not lead to any measurable excitation to other bands.  This scheme was used for all measurements with the exception of the points with $\varphi < 1.5\pi$ in Fig.~2a of the main text, where a single laser was used with one ramp for $\varphi\leq \pi$ and two ramps for $\varphi > \pi$, with a duration of 100\,ms for each ramp.

%%%%%%%% Models for finite pumping efficiency %%%%%%%%%
\section{Models for finite pumping efficiency}

When transitions to other bands occur during the pumping process, the deflection of the cloud can be changed due to the different Chern numbers of the bands. For the measurements in the Wannier tunneling limit, we model the band occupations in a simple two-band model assuming that a small fraction of atoms is excited non-adiabatically to the other band during each cycle. In the two-band model, these transitions will predominantly take place when ramping over the symmetric double well configuration where the gap between the two lowest bands is smallest and atoms tunnel from one site to the next. This situation occurs twice within one pump cycle and we define the transfer efficiency as the fraction of atoms staying in the initially populated band during one ramp over a symmetric double well, which corresponds to half a pump cycle and thus one step in the COM position.

The band populations after the $m$-th step can be expressed as 

\begin{equation}
\boldsymbol{n}^{(m)}=\boldsymbol{\epsilon}^{m}\cdot \boldsymbol{n}^{(0)} 
\label{eqS:BandOcc}
\end{equation}
with
\begin{eqnarray*}
\boldsymbol{n}^{(m)}=\begin{pmatrix} n_1^{(m)} \\ n_2^{(m)} \end{pmatrix}\text{,}\quad 
\boldsymbol{\epsilon}=\begin{pmatrix} 1 - \epsilon_1 & \epsilon_2 \\ \epsilon_1 & 1 - \epsilon_2 \end{pmatrix}
\end{eqnarray*}

\noindent Here, $n_1$ ($n_2$) denotes the fraction of atoms in the first (second) band, $\epsilon_1$ is the fraction of atoms excited from the first to the second band during one step and $\epsilon_2$ the one transferred from the second to the first. While the contribution from non-adiabatic transitions induced by the pumping is expected to be the same for $\epsilon_1$ and $\epsilon_2$, the finite lifetime of atoms in the second band generally leads to $\epsilon_2 > \epsilon_1$. 

When both $\epsilon_1$ and $\epsilon_2$ are very small, one can expand $\boldsymbol{\epsilon}^{m}$ in powers of $\epsilon_1$ and $\epsilon_2$ and obtains to first order
\begin{equation}
\boldsymbol{\epsilon}^m
\approx 
\begin{pmatrix}
1 - m \epsilon_1 & m \epsilon_2 \\
m \epsilon_1 & 1 - m \epsilon_2 \\
\end{pmatrix}
\end{equation}

\noindent Assuming a perfect preparation of the atoms in one band, i.e. $\boldsymbol{n}^{(0)}=(1,0)$ or $\boldsymbol{n}^{(0)}=(0,1)$, the resulting band occupations depend only on $\epsilon_1$ and $\epsilon_2$, respectively. This model was used to fit both the measured COM position as well as the site-resolved band mapping data with a single free parameter.

The expected displacement for both bands can be calculated by evaluating the COM evolution of the corresponding Wannier functions for one half of a pump cycle from $\varphi = 0$ to $\pi$. For each segment $\varphi \in \left](m-1) \pi, m \pi \right]$, the combined motion is obtained by adding up the contributions from the two bands weighted with the respective population fraction $n_1^{(m)}$ and $n_2^{(m)}$. After each half of a pump cycle ($\varphi =m \pi$), the COM position can be expressed in terms of the Chern numbers of the bands:

\begin{align}
x^{(m)}= \frac{d_l}{2}~\boldsymbol\nu \cdot \left(\sum\limits_{j=1}^{m} \mathrm{sign}(m) \boldsymbol{\epsilon}^{|j|} \boldsymbol{n}^{(0)}\right)
\label{eqS:TotalDisplacement}
\end{align}%
\noindent with $\boldsymbol{\nu}=(\nu_1,\nu_2)$. 

For the site occupations, the even-odd distribution was determined by integrating the probability density of the Wannier functions over the even and odd site of the corresponding double well. These were then weighted in the same way with $n_1^{(m)}$ and $n_2^{(m)}$ to obtain the fractions on even and odd sites as a function of $\epsilon_{1}$ and $\epsilon_{2}$.

\begin{figure}[t!]
\includegraphics[width=\linewidth]{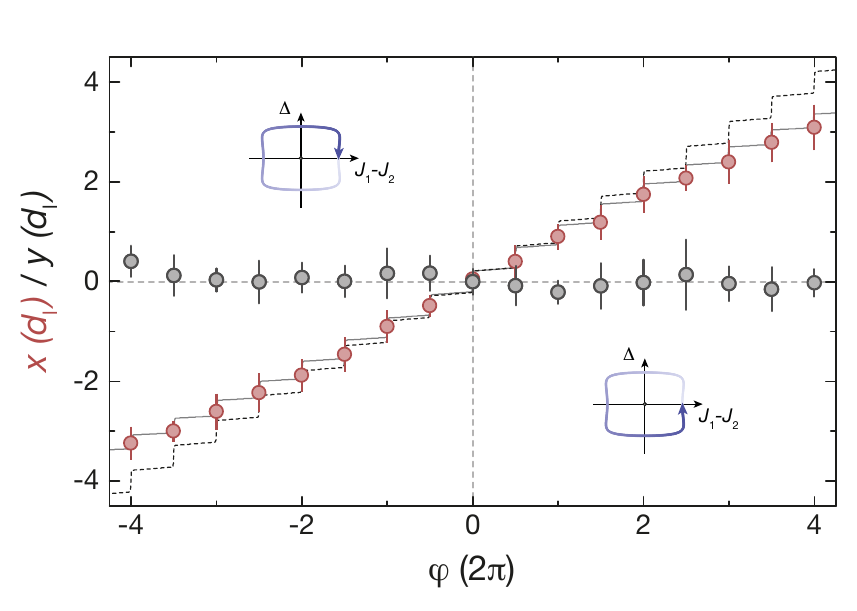}
\vspace{-0.5cm}  
\caption{Evolution of the COM position along the $x$- and $y$-axis when pumping is performed along the $x$-axis. The grey points depict the COM displacement in the $y$-direction for the data points of Fig.2b in the main text, which are illustrated in red. The positions along $y$ were evaluated using the same procedure as for the ones along $x$. Each point is averaged over ten data sets and the error bars depict the standard deviation.}
\label{fig:S3}
\end{figure}

%%%%%%%% Center-of-mass position in the perpendicular direction %%%%%%%%
\section{Center-of-mass position in the perpendicular direction}

In addition to determining the COM position along the pumping direction, we also verified that the pumping along $x$ does not lead to any measurable displacement of the cloud along the perpendicular $y$-direction. For the measurement depicted in Fig.~2b of the main text, the COM positions along both axes are shown in \fig{fig:S3}. As expected, we do not observe any significant displacement of the cloud along $y$ over multiple cycles in both pumping directions.

%%%%%%%% Bibliography %%%%%%%%
\bibliographystyle{bosons}

\begin{thebibliography}{47}%
\makeatletter
\providecommand \@ifxundefined [1]{%
 \@ifx{#1\undefined}
}%
\providecommand \@ifnum [1]{%
 \ifnum #1\expandafter \@firstoftwo
 \else \expandafter \@secondoftwo
 \fi
}%
\providecommand \@ifx [1]{%
 \ifx #1\expandafter \@firstoftwo
 \else \expandafter \@secondoftwo
 \fi
}%
\providecommand \natexlab [1]{#1}%
\providecommand \enquote  [1]{``#1''}%
\providecommand \bibnamefont  [1]{#1}%
\providecommand \bibfnamefont [1]{#1}%
\providecommand \citenamefont [1]{#1}%
\providecommand \href@noop [0]{\@secondoftwo}%
\providecommand \href [0]{\begingroup \@sanitize@url \@href}%
\providecommand \@href[1]{\@@startlink{#1}\@@href}%
\providecommand \@@href[1]{\endgroup#1\@@endlink}%
\providecommand \@sanitize@url [0]{\catcode `\\12\catcode `\$12\catcode
  `\&12\catcode `\#12\catcode `\^12\catcode `\_12\catcode `\%12\relax}%
\providecommand \@@startlink[1]{}%
\providecommand \@@endlink[0]{}%
\providecommand \url  [0]{\begingroup\@sanitize@url \@url }%
\providecommand \@url [1]{\endgroup\@href {#1}{\urlprefix }}%
\providecommand \urlprefix  [0]{URL }%
\providecommand \Eprint [0]{\href }%
\providecommand \doibase [0]{http://dx.doi.org/}%
\providecommand \selectlanguage [0]{\@gobble}%
\providecommand \bibinfo  [0]{\@secondoftwo}%
\providecommand \bibfield  [0]{\@secondoftwo}%
\providecommand \translation [1]{[#1]}%
\providecommand \BibitemOpen [0]{}%
\providecommand \bibitemStop [0]{}%
\providecommand \bibitemNoStop [0]{.\EOS\space}%
\providecommand \EOS [0]{\spacefactor3000\relax}%
\providecommand \BibitemShut  [1]{\csname bibitem#1\endcsname}%
\let\auto@bib@innerbib\@empty
%</preamble>
\bibitem [{\citenamefont {Thouless}(1983)}]{Thouless:1983}%
  \BibitemOpen
  \bibfield  {author} {\bibinfo {author} {\bibfnamefont {D.~J.}\ \bibnamefont
  {Thouless}},\ }\href {\doibase 10.1103/PhysRevB.27.6083} {\bibfield
  {journal} {\bibinfo  {journal} {Phys. Rev. B}\ }\textbf {\bibinfo {volume}
  {27}},\ \bibinfo {pages} {6083}} (\bibinfo {year} {1983})\BibitemShut
  {NoStop}%
\bibitem [{\citenamefont {Niu\ and\ Thouless}(1984)\citenamefont {Niu}\ and\
  \citenamefont {Thouless}}]{Niu:1984}%
  \BibitemOpen
  \bibfield  {author} {\bibinfo {author} {\bibfnamefont {Q.}~\bibnamefont
  {Niu}}\ \bibnamefont {and}\ \bibinfo {author} {\bibfnamefont {D.~J.}\
  \bibnamefont {Thouless}},\ }\href {\doibase doi:10.1088/0305-4470/17/12/016}
  {\bibfield  {journal} {\bibinfo  {journal} {J. Phys. A}\ }\textbf {\bibinfo
  {volume} {17}},\ \bibinfo {pages} {2453}} (\bibinfo {year}
  {1984})\BibitemShut {NoStop}%
\bibitem [{\citenamefont {Klitzing et~al.}(1980)\citenamefont {Klitzing},
  \citenamefont {Dorda},\ and\ \citenamefont {Pepper}}]{Klitzing:1980}%
  \BibitemOpen
  \bibfield  {author} {\bibinfo {author} {\bibfnamefont {K.~V.}\ \bibnamefont
  {Klitzing}}, \bibinfo {author} {\bibfnamefont {G.}~\bibnamefont {Dorda}},\
  \bibnamefont {and}\ \bibinfo {author} {\bibfnamefont {M.}~\bibnamefont
  {Pepper}},\ }\href {\doibase 10.1103/PhysRevLett.45.494} {\bibfield
  {journal} {\bibinfo  {journal} {Phys. Rev. Lett.}\ }\textbf {\bibinfo
  {volume} {45}},\ \bibinfo {pages} {494}} (\bibinfo {year} {1980})\BibitemShut
  {NoStop}%
\bibitem [{\citenamefont {Thouless et~al.}(1982)\citenamefont {Thouless},
  \citenamefont {Kohmoto}, \citenamefont {Nightingale},\ and\ \citenamefont
  {{Den Nijs}}}]{Thouless:1982}%
  \BibitemOpen
  \bibfield  {author} {\bibinfo {author} {\bibfnamefont {D.}~\bibnamefont
  {Thouless}}, \bibinfo {author} {\bibfnamefont {M.}~\bibnamefont {Kohmoto}},
  \bibinfo {author} {\bibfnamefont {M.}~\bibnamefont {Nightingale}},\
  \bibnamefont {and}\ \bibinfo {author} {\bibfnamefont {M.}~\bibnamefont {{Den
  Nijs}}},\ }\href {\doibase 10.1103/PhysRevLett.49.405} {\bibfield  {journal}
  {\bibinfo  {journal} {Phys. Rev. Lett.}\ }\textbf {\bibinfo {volume} {49}},\
  \bibinfo {pages} {405}} (\bibinfo {year} {1982})\BibitemShut {NoStop}%
\bibitem [{\citenamefont {Niu}(1990)}]{Niu:1990}%
  \BibitemOpen
  \bibfield  {author} {\bibinfo {author} {\bibfnamefont {Q.}~\bibnamefont
  {Niu}},\ }\href {\doibase 10.1103/PhysRevLett.64.1812} {\bibfield  {journal}
  {\bibinfo  {journal} {Phys. Rev. Lett.}\ }\textbf {\bibinfo {volume} {64}},\
  \bibinfo {pages} {1812}} (\bibinfo {year} {1990})\BibitemShut {NoStop}%
\bibitem [{\citenamefont {Pekola et~al.}(2013)\citenamefont {Pekola},
  \citenamefont {Saira}, \citenamefont {Maisi}, \citenamefont {Kemppinen},
  \citenamefont {M\"ott\"onen}, \citenamefont {Pashkin},\ and\ \citenamefont
  {Averin}}]{Pekola:2013}%
  \BibitemOpen
  \bibfield  {author} {\bibinfo {author} {\bibfnamefont {J.~P.}\ \bibnamefont
  {Pekola}}, \bibinfo {author} {\bibfnamefont {O.-P.}\ \bibnamefont {Saira}},
  \bibinfo {author} {\bibfnamefont {V.~F.}\ \bibnamefont {Maisi}}, \bibnamefont
  {et~al.},\ }\href {\doibase 10.1103/RevModPhys.85.1421} {\bibfield  {journal}
  {\bibinfo  {journal} {Rev. Mod. Phys.}\ }\textbf {\bibinfo {volume} {85}},\
  \bibinfo {pages} {1421}} (\bibinfo {year} {2013})\BibitemShut {NoStop}%
\bibitem [{\citenamefont {Splettstoesser et~al.}(2005)\citenamefont
  {Splettstoesser}, \citenamefont {Governale}, \citenamefont {K\"{o}nig},\ and\
  \citenamefont {Fazio}}]{Splettstoesser:2005}%
  \BibitemOpen
  \bibfield  {author} {\bibinfo {author} {\bibfnamefont {J.}~\bibnamefont
  {Splettstoesser}}, \bibinfo {author} {\bibfnamefont {M.}~\bibnamefont
  {Governale}}, \bibinfo {author} {\bibfnamefont {J.}~\bibnamefont
  {K\"{o}nig}},\ \bibnamefont {and}\ \bibinfo {author} {\bibfnamefont
  {R.}~\bibnamefont {Fazio}},\ }\href {\doibase 10.1103/PhysRevLett.95.246803}
  {\bibfield  {journal} {\bibinfo  {journal} {Phys. Rev. Lett.}\ }\textbf
  {\bibinfo {volume} {95}},\ \bibinfo {pages} {3}} (\bibinfo {year}
  {2005})\BibitemShut {NoStop}%
\bibitem [{\citenamefont {Marra et~al.}(2015)\citenamefont {Marra},
  \citenamefont {Citro},\ and\ \citenamefont {Ortix}}]{Marra:2015}%
  \BibitemOpen
  \bibfield  {author} {\bibinfo {author} {\bibfnamefont {P.}~\bibnamefont
  {Marra}}, \bibinfo {author} {\bibfnamefont {R.}~\bibnamefont {Citro}},\
  \bibnamefont {and}\ \bibinfo {author} {\bibfnamefont {C.}~\bibnamefont
  {Ortix}},\ }\href {\doibase 10.1103/PhysRevB.91.125411} {\bibfield  {journal}
  {\bibinfo  {journal} {Phys. Rev. B}\ }\textbf {\bibinfo {volume} {91}},\
  \bibinfo {pages} {2}} (\bibinfo {year} {2015})\BibitemShut {NoStop}%
\bibitem [{\citenamefont {Pothier et~al.}(1992)\citenamefont {Pothier},
  \citenamefont {Lafarge}, \citenamefont {Urbina}, \citenamefont {Esteve},\
  and\ \citenamefont {Devoret}}]{Pothier:1992}%
  \BibitemOpen
  \bibfield  {author} {\bibinfo {author} {\bibfnamefont {H.}~\bibnamefont
  {Pothier}}, \bibinfo {author} {\bibfnamefont {P.}~\bibnamefont {Lafarge}},
  \bibinfo {author} {\bibfnamefont {C.}~\bibnamefont {Urbina}}, \bibinfo
  {author} {\bibfnamefont {D.}~\bibnamefont {Esteve}},\ \bibnamefont {and}\
  \bibinfo {author} {\bibfnamefont {M.~H.}\ \bibnamefont {Devoret}},\ }\href
  {http://stacks.iop.org/0295-5075/17/i=3/a=011} {\bibfield  {journal}
  {\bibinfo  {journal} {EPL}\ }\textbf {\bibinfo {volume} {17}},\ \bibinfo
  {pages} {249}} (\bibinfo {year} {1992})\BibitemShut {NoStop}%
\bibitem [{\citenamefont {Talyanskii et~al.}(1997)\citenamefont {Talyanskii},
  \citenamefont {Shilton}, \citenamefont {Pepper}, \citenamefont {Smith},
  \citenamefont {Ford}, \citenamefont {Linfield}, \citenamefont {Ritchie},\
  and\ \citenamefont {Jones}}]{Talyanskii:1997}%
  \BibitemOpen
  \bibfield  {author} {\bibinfo {author} {\bibfnamefont {V.~I.}\ \bibnamefont
  {Talyanskii}}, \bibinfo {author} {\bibfnamefont {J.~M.}\ \bibnamefont
  {Shilton}}, \bibinfo {author} {\bibfnamefont {M.}~\bibnamefont {Pepper}},
  \bibnamefont {et~al.},\ }\href {\doibase 10.1103/PhysRevB.56.15180}
  {\bibfield  {journal} {\bibinfo  {journal} {Phys. Rev. B}\ }\textbf {\bibinfo
  {volume} {56}},\ \bibinfo {pages} {15180}} (\bibinfo {year}
  {1997})\BibitemShut {NoStop}%
\bibitem [{\citenamefont {Blumenthal et~al.}(2007)\citenamefont {Blumenthal},
  \citenamefont {Kaestner}, \citenamefont {Li}, \citenamefont {Giblin},
  \citenamefont {Janssen}, \citenamefont {Pepper}, \citenamefont {Anderson},
  \citenamefont {Jones},\ and\ \citenamefont {Ritchie}}]{Blumenthal:2007}%
  \BibitemOpen
  \bibfield  {author} {\bibinfo {author} {\bibfnamefont {M.~D.}\ \bibnamefont
  {Blumenthal}}, \bibinfo {author} {\bibfnamefont {B.}~\bibnamefont
  {Kaestner}}, \bibinfo {author} {\bibfnamefont {L.}~\bibnamefont {Li}},
  \bibnamefont {et~al.},\ }\href {\doibase 10.1038/nphys582} {\bibfield
  {journal} {\bibinfo  {journal} {Nature Phys.}\ }\textbf {\bibinfo {volume}
  {3}},\ \bibinfo {pages} {343}} (\bibinfo {year} {2007})\BibitemShut {NoStop}%
\bibitem [{\citenamefont {Switkes et~al.}(1999)\citenamefont {Switkes},
  \citenamefont {Marcus}, \citenamefont {Campman},\ and\ \citenamefont
  {Gossard}}]{Switkes:1999}%
  \BibitemOpen
  \bibfield  {author} {\bibinfo {author} {\bibfnamefont {M.}~\bibnamefont
  {Switkes}}, \bibinfo {author} {\bibfnamefont {C.~M.}\ \bibnamefont {Marcus}},
  \bibinfo {author} {\bibfnamefont {K.}~\bibnamefont {Campman}},\ \bibnamefont
  {and}\ \bibinfo {author} {\bibfnamefont {A.~C.}\ \bibnamefont {Gossard}},\
  }\href {\doibase 10.1126/science.283.5409.1905} {\bibfield  {journal}
  {\bibinfo  {journal} {Science}\ }\textbf {\bibinfo {volume} {283}},\ \bibinfo
  {pages} {10}} (\bibinfo {year} {1999})\BibitemShut {NoStop}%
\bibitem [{\citenamefont {Brouwer}(1998)}]{Brouwer:1998}%
  \BibitemOpen
  \bibfield  {author} {\bibinfo {author} {\bibfnamefont {P.~W.}\ \bibnamefont
  {Brouwer}},\ }\href {\doibase 10.1103/PhysRevB.58.R10135} {\bibfield
  {journal} {\bibinfo  {journal} {Phys. Rev. B}\ }\textbf {\bibinfo {volume}
  {58}},\ \bibinfo {pages} {R10135}} (\bibinfo {year} {1998})\BibitemShut
  {NoStop}%
\bibitem [{\citenamefont {Kraus et~al.}(2012)\citenamefont {Kraus},
  \citenamefont {Lahini}, \citenamefont {Ringel}, \citenamefont {Verbin},\ and\
  \citenamefont {Zilberberg}}]{Kraus:2012}%
  \BibitemOpen
  \bibfield  {author} {\bibinfo {author} {\bibfnamefont {Y.~E.}\ \bibnamefont
  {Kraus}}, \bibinfo {author} {\bibfnamefont {Y.}~\bibnamefont {Lahini}},
  \bibinfo {author} {\bibfnamefont {Z.}~\bibnamefont {Ringel}}, \bibinfo
  {author} {\bibfnamefont {M.}~\bibnamefont {Verbin}},\ \bibnamefont {and}\
  \bibinfo {author} {\bibfnamefont {O.}~\bibnamefont {Zilberberg}},\ }\href
  {\doibase 10.1103/PhysRevLett.109.106402} {\bibfield  {journal} {\bibinfo
  {journal} {Phys. Rev. Lett.}\ }\textbf {\bibinfo {volume} {109}},\ \bibinfo
  {pages} {106402}} (\bibinfo {year} {2012})\BibitemShut {NoStop}%
\bibitem [{\citenamefont {Verbin et~al.}(2015)\citenamefont {Verbin},
  \citenamefont {Zilberberg}, \citenamefont {Lahini}, \citenamefont {Kraus},\
  and\ \citenamefont {Silberberg}}]{Verbin:2015}%
  \BibitemOpen
  \bibfield  {author} {\bibinfo {author} {\bibfnamefont {M.}~\bibnamefont
  {Verbin}}, \bibinfo {author} {\bibfnamefont {O.}~\bibnamefont {Zilberberg}},
  \bibinfo {author} {\bibfnamefont {Y.}~\bibnamefont {Lahini}}, \bibinfo
  {author} {\bibfnamefont {Y.~E.}\ \bibnamefont {Kraus}},\ \bibnamefont {and}\
  \bibinfo {author} {\bibfnamefont {Y.}~\bibnamefont {Silberberg}},\ }\href
  {\doibase 10.1103/PhysRevB.91.064201} {\bibfield  {journal} {\bibinfo
  {journal} {Phys. Rev. B}\ }\textbf {\bibinfo {volume} {91}},\ \bibinfo
  {pages} {064201}} (\bibinfo {year} {2015})\BibitemShut {NoStop}%
\bibitem [{\citenamefont {Atala et~al.}(2013)\citenamefont {Atala},
  \citenamefont {Aidelsburger}, \citenamefont {Barreiro}, \citenamefont
  {Abanin}, \citenamefont {Kitagawa}, \citenamefont {Demler},\ and\
  \citenamefont {Bloch}}]{Atala:2013}%
  \BibitemOpen
  \bibfield  {author} {\bibinfo {author} {\bibfnamefont {M.}~\bibnamefont
  {Atala}}, \bibinfo {author} {\bibfnamefont {M.}~\bibnamefont {Aidelsburger}},
  \bibinfo {author} {\bibfnamefont {J.~T.}\ \bibnamefont {Barreiro}},
  \bibnamefont {et~al.},\ }\href {http://dx.doi.org/10.1038/nphys2790}
  {\bibfield  {journal} {\bibinfo  {journal} {Nature Phys.}\ }\textbf {\bibinfo
  {volume} {9}},\ \bibinfo {pages} {795}} (\bibinfo {year} {2013})\BibitemShut
  {NoStop}%
\bibitem [{\citenamefont {Aidelsburger et~al.}(2013)\citenamefont
  {Aidelsburger}, \citenamefont {Atala}, \citenamefont {Lohse}, \citenamefont
  {Barreiro}, \citenamefont {Paredes},\ and\ \citenamefont
  {Bloch}}]{Aidelsburger:2013}%
  \BibitemOpen
  \bibfield  {author} {\bibinfo {author} {\bibfnamefont {M.}~\bibnamefont
  {Aidelsburger}}, \bibinfo {author} {\bibfnamefont {M.}~\bibnamefont {Atala}},
  \bibinfo {author} {\bibfnamefont {M.}~\bibnamefont {Lohse}}, \bibnamefont
  {et~al.},\ }\href {\doibase 10.1103/PhysRevLett.111.185301} {\bibfield
  {journal} {\bibinfo  {journal} {Phys. Rev. Lett.}\ }\textbf {\bibinfo
  {volume} {111}},\ \bibinfo {pages} {185301}} (\bibinfo {year}
  {2013})\BibitemShut {NoStop}%
\bibitem [{\citenamefont {Miyake et~al.}(2013)\citenamefont {Miyake},
  \citenamefont {Siviloglou}, \citenamefont {Kennedy}, \citenamefont {Burton},\
  and\ \citenamefont {Ketterle}}]{Miyake:2013}%
  \BibitemOpen
  \bibfield  {author} {\bibinfo {author} {\bibfnamefont {H.}~\bibnamefont
  {Miyake}}, \bibinfo {author} {\bibfnamefont {G.~A.}\ \bibnamefont
  {Siviloglou}}, \bibinfo {author} {\bibfnamefont {C.~J.}\ \bibnamefont
  {Kennedy}}, \bibinfo {author} {\bibfnamefont {W.~C.}\ \bibnamefont
  {Burton}},\ \bibnamefont {and}\ \bibinfo {author} {\bibfnamefont
  {W.}~\bibnamefont {Ketterle}},\ }\href {\doibase
  10.1103/PhysRevLett.111.185302} {\bibfield  {journal} {\bibinfo  {journal}
  {Phys. Rev. Lett.}\ }\textbf {\bibinfo {volume} {111}},\ \bibinfo {pages}
  {185302}} (\bibinfo {year} {2013})\BibitemShut {NoStop}%
\bibitem [{\citenamefont {{Mancini} et~al.}(2015)\citenamefont {{Mancini}},
  \citenamefont {{Pagano}}, \citenamefont {{Cappellini}}, \citenamefont
  {{Livi}}, \citenamefont {{Rider}}, \citenamefont {{Catani}}, \citenamefont
  {{Sias}}, \citenamefont {{Zoller}}, \citenamefont {{Inguscio}}, \citenamefont
  {{Dalmonte}},\ and\ \citenamefont {{Fallani}}}]{Mancini:2015}%
  \BibitemOpen
  \bibfield  {author} {\bibinfo {author} {\bibfnamefont {M.}~\bibnamefont
  {{Mancini}}}, \bibinfo {author} {\bibfnamefont {G.}~\bibnamefont {{Pagano}}},
  \bibinfo {author} {\bibfnamefont {G.}~\bibnamefont {{Cappellini}}},
  \bibnamefont {et~al.},\ }\href@noop {} {\Eprint
  {http://arxiv.org/abs/1502.02495} {arXiv:1502.02495 [cond-mat.quant-gas]} }
  (\bibinfo {year} {2015})\BibitemShut {NoStop}%
\bibitem [{\citenamefont {{Stuhl} et~al.}(2015)\citenamefont {{Stuhl}},
  \citenamefont {{Lu}}, \citenamefont {{Aycock}}, \citenamefont {{Genkina}},\
  and\ \citenamefont {{Spielman}}}]{Stuhl:2015}%
  \BibitemOpen
  \bibfield  {author} {\bibinfo {author} {\bibfnamefont {B.~K.}\ \bibnamefont
  {{Stuhl}}}, \bibinfo {author} {\bibfnamefont {H.-I.}\ \bibnamefont {{Lu}}},
  \bibinfo {author} {\bibfnamefont {L.~M.}\ \bibnamefont {{Aycock}}}, \bibinfo
  {author} {\bibfnamefont {D.}~\bibnamefont {{Genkina}}},\ \bibnamefont {and}\
  \bibinfo {author} {\bibfnamefont {I.~B.}\ \bibnamefont {{Spielman}}},\
  }\href@noop {} {\Eprint {http://arxiv.org/abs/1502.02496} {arXiv:1502.02496
  [cond-mat.quant-gas]} } (\bibinfo {year} {2015})\BibitemShut {NoStop}%
\bibitem [{\citenamefont {Jotzu et~al.}(2014)\citenamefont {Jotzu},
  \citenamefont {Messer}, \citenamefont {Desbuquois}, \citenamefont {Lebrat},
  \citenamefont {Uehlinger}, \citenamefont {Greif},\ and\ \citenamefont
  {Esslinger}}]{Jotzu:2014}%
  \BibitemOpen
  \bibfield  {author} {\bibinfo {author} {\bibfnamefont {G.}~\bibnamefont
  {Jotzu}}, \bibinfo {author} {\bibfnamefont {M.}~\bibnamefont {Messer}},
  \bibinfo {author} {\bibfnamefont {R.}~\bibnamefont {Desbuquois}},
  \bibnamefont {et~al.},\ }\href {http://dx.doi.org/10.1038/nature13915}
  {\bibfield  {journal} {\bibinfo  {journal} {Nature}\ }\textbf {\bibinfo
  {volume} {515}},\ \bibinfo {pages} {237}} (\bibinfo {year}
  {2014})\BibitemShut {NoStop}%
\bibitem [{\citenamefont {Duca et~al.}(2015)\citenamefont {Duca}, \citenamefont
  {Li}, \citenamefont {Reitter}, \citenamefont {Bloch}, \citenamefont
  {Schleier-Smith},\ and\ \citenamefont {Schneider}}]{Duca:2015}%
  \BibitemOpen
  \bibfield  {author} {\bibinfo {author} {\bibfnamefont {L.}~\bibnamefont
  {Duca}}, \bibinfo {author} {\bibfnamefont {T.}~\bibnamefont {Li}}, \bibinfo
  {author} {\bibfnamefont {M.}~\bibnamefont {Reitter}}, \bibnamefont {et~al.},\
  }\href {\doibase 10.1126/science.1259052} {\bibfield  {journal} {\bibinfo
  {journal} {Science}\ }\textbf {\bibinfo {volume} {347}},\ \bibinfo {pages}
  {288}} (\bibinfo {year} {2015})\BibitemShut {NoStop}%
\bibitem [{\citenamefont {Aidelsburger et~al.}(2015)\citenamefont
  {Aidelsburger}, \citenamefont {Lohse}, \citenamefont {Schweizer},
  \citenamefont {Atala}, \citenamefont {Barreiro}, \citenamefont {Nascimbene},
  \citenamefont {Cooper}, \citenamefont {Bloch},\ and\ \citenamefont
  {Goldman}}]{Aidelsburger:2015}%
  \BibitemOpen
  \bibfield  {author} {\bibinfo {author} {\bibfnamefont {M.}~\bibnamefont
  {Aidelsburger}}, \bibinfo {author} {\bibfnamefont {M.}~\bibnamefont {Lohse}},
  \bibinfo {author} {\bibfnamefont {C.}~\bibnamefont {Schweizer}}, \bibnamefont
  {et~al.},\ }\href {http://dx.doi.org/10.1038/nphys3171} {\bibfield  {journal}
  {\bibinfo  {journal} {Nature Phys.}\ }\textbf {\bibinfo {volume} {11}},\
  \bibinfo {pages} {162}} (\bibinfo {year} {2015})\BibitemShut {NoStop}%
\bibitem [{\citenamefont {Romero-Isart\ and\
  Garc\'{\i}a-Ripoll}(2007)\citenamefont {Romero-Isart}\ and\ \citenamefont
  {Garc\'{\i}a-Ripoll}}]{Romero-Isart:2007}%
  \BibitemOpen
  \bibfield  {author} {\bibinfo {author} {\bibfnamefont {O.}~\bibnamefont
  {Romero-Isart}}\ \bibnamefont {and}\ \bibinfo {author} {\bibfnamefont
  {J.~J.}\ \bibnamefont {Garc\'{\i}a-Ripoll}},\ }\href {\doibase
  10.1103/PhysRevA.76.052304} {\bibfield  {journal} {\bibinfo  {journal} {Phys.
  Rev. A}\ }\textbf {\bibinfo {volume} {76}},\ \bibinfo {pages} {052304}}
  (\bibinfo {year} {2007})\BibitemShut {NoStop}%
\bibitem [{\citenamefont {Qian et~al.}(2011)\citenamefont {Qian}, \citenamefont
  {Gong},\ and\ \citenamefont {Zhang}}]{Qian:2011}%
  \BibitemOpen
  \bibfield  {author} {\bibinfo {author} {\bibfnamefont {Y.}~\bibnamefont
  {Qian}}, \bibinfo {author} {\bibfnamefont {M.}~\bibnamefont {Gong}},\
  \bibnamefont {and}\ \bibinfo {author} {\bibfnamefont {C.}~\bibnamefont
  {Zhang}},\ }\href {\doibase 10.1103/PhysRevA.84.013608} {\bibfield  {journal}
  {\bibinfo  {journal} {Phys. Rev. A}\ }\textbf {\bibinfo {volume} {84}},\
  \bibinfo {pages} {013608}} (\bibinfo {year} {2011})\BibitemShut {NoStop}%
\bibitem [{\citenamefont {Wang et~al.}(2013)\citenamefont {Wang}, \citenamefont
  {Troyer},\ and\ \citenamefont {Dai}}]{Wang:2013}%
  \BibitemOpen
  \bibfield  {author} {\bibinfo {author} {\bibfnamefont {L.}~\bibnamefont
  {Wang}}, \bibinfo {author} {\bibfnamefont {M.}~\bibnamefont {Troyer}},\
  \bibnamefont {and}\ \bibinfo {author} {\bibfnamefont {X.}~\bibnamefont
  {Dai}},\ }\href {\doibase 10.1103/PhysRevLett.111.026802} {\bibfield
  {journal} {\bibinfo  {journal} {Phys. Rev. Lett.}\ }\textbf {\bibinfo
  {volume} {111}},\ \bibinfo {pages} {026802}} (\bibinfo {year}
  {2013})\BibitemShut {NoStop}%
\bibitem [{\citenamefont {Karplus\ and\ Luttinger}(1954)\citenamefont
  {Karplus}\ and\ \citenamefont {Luttinger}}]{Karplus:1954}%
  \BibitemOpen
  \bibfield  {author} {\bibinfo {author} {\bibfnamefont {R.}~\bibnamefont
  {Karplus}}\ \bibnamefont {and}\ \bibinfo {author} {\bibfnamefont {J.~M.}\
  \bibnamefont {Luttinger}},\ }\href {\doibase 10.1103/PhysRev.95.1154}
  {\bibfield  {journal} {\bibinfo  {journal} {Phys. Rev.}\ }\textbf {\bibinfo
  {volume} {95}},\ \bibinfo {pages} {1154}} (\bibinfo {year}
  {1954})\BibitemShut {NoStop}%
\bibitem [{\citenamefont {Xiao et~al.}(2010)\citenamefont {Xiao}, \citenamefont
  {Chang},\ and\ \citenamefont {Niu}}]{Xiao:2010}%
  \BibitemOpen
  \bibfield  {author} {\bibinfo {author} {\bibfnamefont {D.}~\bibnamefont
  {Xiao}}, \bibinfo {author} {\bibfnamefont {M.-C.}\ \bibnamefont {Chang}},\
  \bibnamefont {and}\ \bibinfo {author} {\bibfnamefont {Q.}~\bibnamefont
  {Niu}},\ }\href {\doibase 10.1103/RevModPhys.82.1959} {\bibfield  {journal}
  {\bibinfo  {journal} {Rev. Mod. Phys.}\ }\textbf {\bibinfo {volume} {82}},\
  \bibinfo {pages} {1959}} (\bibinfo {year} {2010})\BibitemShut {NoStop}%
\bibitem [{\citenamefont {{Wei}\ and\ {Mueller}}(2015)\citenamefont {{Wei}}\
  and\ \citenamefont {{Mueller}}}]{Wei:2015}%
  \BibitemOpen
  \bibfield  {author} {\bibinfo {author} {\bibfnamefont {R.}~\bibnamefont
  {{Wei}}}\ \bibnamefont {and}\ \bibinfo {author} {\bibfnamefont {E.~J.}\
  \bibnamefont {{Mueller}}},\ }\href@noop {} {\Eprint
  {http://arxiv.org/abs/1502.04208} {arXiv:1502.04208 [cond-mat.quant-gas]} }
  (\bibinfo {year} {2015})\BibitemShut {NoStop}%
\bibitem [{\citenamefont {Laughlin}(1981)}]{Laughlin:1981}%
  \BibitemOpen
  \bibfield  {author} {\bibinfo {author} {\bibfnamefont {R.~B.}\ \bibnamefont
  {Laughlin}},\ }\href {\doibase 10.1103/PhysRevB.23.5632} {\bibfield
  {journal} {\bibinfo  {journal} {Phys. Rev. B}\ }\textbf {\bibinfo {volume}
  {23}},\ \bibinfo {pages} {5632}} (\bibinfo {year} {1981})\BibitemShut
  {NoStop}%
\bibitem [{\citenamefont {Harper}(1955)}]{Harper:1955}%
  \BibitemOpen
  \bibfield  {author} {\bibinfo {author} {\bibfnamefont {P.~G.}\ \bibnamefont
  {Harper}},\ }\href {http://stacks.iop.org/0370-1298/68/i=10/a=305} {\bibfield
   {journal} {\bibinfo  {journal} {Proc. Phys. Soc. A}\ }\textbf {\bibinfo
  {volume} {68}},\ \bibinfo {pages} {879}} (\bibinfo {year} {1955})\BibitemShut
  {NoStop}%
\bibitem [{\citenamefont {Roux et~al.}(2008)\citenamefont {Roux}, \citenamefont
  {Barthel}, \citenamefont {McCulloch}, \citenamefont {Kollath}, \citenamefont
  {Schollw\"ock},\ and\ \citenamefont {Giamarchi}}]{Roux:2008}%
  \BibitemOpen
  \bibfield  {author} {\bibinfo {author} {\bibfnamefont {G.}~\bibnamefont
  {Roux}}, \bibinfo {author} {\bibfnamefont {T.}~\bibnamefont {Barthel}},
  \bibinfo {author} {\bibfnamefont {I.~P.}\ \bibnamefont {McCulloch}},
  \bibnamefont {et~al.},\ }\href {\doibase 10.1103/PhysRevA.78.023628}
  {\bibfield  {journal} {\bibinfo  {journal} {Phys. Rev. A}\ }\textbf {\bibinfo
  {volume} {78}},\ \bibinfo {pages} {023628}} (\bibinfo {year}
  {2008})\BibitemShut {NoStop}%
\bibitem [{\citenamefont {Kraus\ and\ Zilberberg}(2012)\citenamefont {Kraus}\
  and\ \citenamefont {Zilberberg}}]{Kraus:2012a}%
  \BibitemOpen
  \bibfield  {author} {\bibinfo {author} {\bibfnamefont {Y.~E.}\ \bibnamefont
  {Kraus}}\ \bibnamefont {and}\ \bibinfo {author} {\bibfnamefont
  {O.}~\bibnamefont {Zilberberg}},\ }\href {\doibase
  10.1103/PhysRevLett.109.116404} {\bibfield  {journal} {\bibinfo  {journal}
  {Phys. Rev. Lett.}\ }\textbf {\bibinfo {volume} {109}},\ \bibinfo {pages}
  {116404}} (\bibinfo {year} {2012})\BibitemShut {NoStop}%
\bibitem [{\citenamefont {Kraus et~al.}(2013)\citenamefont {Kraus},
  \citenamefont {Ringel},\ and\ \citenamefont {Zilberberg}}]{Kraus:2013}%
  \BibitemOpen
  \bibfield  {author} {\bibinfo {author} {\bibfnamefont {Y.~E.}\ \bibnamefont
  {Kraus}}, \bibinfo {author} {\bibfnamefont {Z.}~\bibnamefont {Ringel}},\
  \bibnamefont {and}\ \bibinfo {author} {\bibfnamefont {O.}~\bibnamefont
  {Zilberberg}},\ }\href {\doibase 10.1103/PhysRevLett.111.226401} {\bibfield
  {journal} {\bibinfo  {journal} {Phys. Rev. Lett.}\ }\textbf {\bibinfo
  {volume} {111}},\ \bibinfo {pages} {226401}} (\bibinfo {year}
  {2013})\BibitemShut {NoStop}%
\bibitem [{\citenamefont {Azbel}(1964)}]{Azbel:1964}%
  \BibitemOpen
  \bibfield  {author} {\bibinfo {author} {\bibfnamefont {M.~Y.}\ \bibnamefont
  {Azbel}},\ }\href@noop {} {\bibfield  {journal} {\bibinfo  {journal} {Zh.
  Eksp. Teor. Fiz.}\ }\textbf {\bibinfo {volume} {46}},\ \bibinfo {pages}
  {929}} (\bibinfo {year} {1964}),\ \bibinfo {note} {[Sov. Phys. JETP
  \textbf{19}, 634 (1964)]}\BibitemShut {NoStop}%
\bibitem [{\citenamefont {Hofstadter}(1976)}]{Hofstadter:1976}%
  \BibitemOpen
  \bibfield  {author} {\bibinfo {author} {\bibfnamefont {D.~R.}\ \bibnamefont
  {Hofstadter}},\ }\href {\doibase 10.1103/PhysRevB.14.2239} {\bibfield
  {journal} {\bibinfo  {journal} {Phys. Rev. B}\ }\textbf {\bibinfo {volume}
  {14}},\ \bibinfo {pages} {2239}} (\bibinfo {year} {1976})\BibitemShut
  {NoStop}%
\bibitem [{\citenamefont {Hatsugai\ and\ Kohmoto}(1990)\citenamefont
  {Hatsugai}\ and\ \citenamefont {Kohmoto}}]{Hatsugai:1990}%
  \BibitemOpen
  \bibfield  {author} {\bibinfo {author} {\bibfnamefont {Y.}~\bibnamefont
  {Hatsugai}}\ \bibnamefont {and}\ \bibinfo {author} {\bibfnamefont
  {M.}~\bibnamefont {Kohmoto}},\ }\href {\doibase 10.1103/PhysRevB.42.8282}
  {\bibfield  {journal} {\bibinfo  {journal} {Phys. Rev. B}\ }\textbf {\bibinfo
  {volume} {42}},\ \bibinfo {pages} {8282}} (\bibinfo {year}
  {1990})\BibitemShut {NoStop}%
\bibitem [{\citenamefont {Rice\ and\ Mele}(1982)\citenamefont {Rice}\ and\
  \citenamefont {Mele}}]{Rice:1982}%
  \BibitemOpen
  \bibfield  {author} {\bibinfo {author} {\bibfnamefont {M.~J.}\ \bibnamefont
  {Rice}}\ \bibnamefont {and}\ \bibinfo {author} {\bibfnamefont {E.~J.}\
  \bibnamefont {Mele}},\ }\href {\doibase 10.1103/PhysRevLett.49.1455}
  {\bibfield  {journal} {\bibinfo  {journal} {Phys. Rev. Lett.}\ }\textbf
  {\bibinfo {volume} {49}},\ \bibinfo {pages} {1455}} (\bibinfo {year}
  {1982})\BibitemShut {NoStop}%
\bibitem [{\citenamefont {Su et~al.}(1979)\citenamefont {Su}, \citenamefont
  {Schrieffer},\ and\ \citenamefont {Heeger}}]{Su:1979}%
  \BibitemOpen
  \bibfield  {author} {\bibinfo {author} {\bibfnamefont {W.~P.}\ \bibnamefont
  {Su}}, \bibinfo {author} {\bibfnamefont {J.~R.}\ \bibnamefont {Schrieffer}},\
  \bibnamefont {and}\ \bibinfo {author} {\bibfnamefont {A.~J.}\ \bibnamefont
  {Heeger}},\ }\href {\doibase 10.1103/PhysRevLett.42.1698} {\bibfield
  {journal} {\bibinfo  {journal} {Phys. Rev. Lett.}\ }\textbf {\bibinfo
  {volume} {42}},\ \bibinfo {pages} {1698}} (\bibinfo {year}
  {1979})\BibitemShut {NoStop}%
\bibitem [{\citenamefont {Weitenberg et~al.}(2011)\citenamefont {Weitenberg},
  \citenamefont {Endres}, \citenamefont {Sherson}, \citenamefont {Cheneau},
  \citenamefont {Schau{\ss}}, \citenamefont {Fukuhara}, \citenamefont {Bloch},\
  and\ \citenamefont {Kuhr}}]{Weitenberg:2011}%
  \BibitemOpen
  \bibfield  {author} {\bibinfo {author} {\bibfnamefont {C.}~\bibnamefont
  {Weitenberg}}, \bibinfo {author} {\bibfnamefont {M.}~\bibnamefont {Endres}},
  \bibinfo {author} {\bibfnamefont {J.~F.}\ \bibnamefont {Sherson}},
  \bibnamefont {et~al.},\ }\href {http://dx.doi.org/10.1038/nature09827}
  {\bibfield  {journal} {\bibinfo  {journal} {Nature}\ }\textbf {\bibinfo
  {volume} {471}},\ \bibinfo {pages} {319}} (\bibinfo {year}
  {2011})\BibitemShut {NoStop}%
\bibitem [{\citenamefont {Preiss et~al.}(2015)\citenamefont {Preiss},
  \citenamefont {Ma}, \citenamefont {Tai}, \citenamefont {Lukin}, \citenamefont
  {Rispoli}, \citenamefont {Zupancic}, \citenamefont {Lahini}, \citenamefont
  {Islam},\ and\ \citenamefont {Greiner}}]{Preiss:2015}%
  \BibitemOpen
  \bibfield  {author} {\bibinfo {author} {\bibfnamefont {P.~M.}\ \bibnamefont
  {Preiss}}, \bibinfo {author} {\bibfnamefont {R.}~\bibnamefont {Ma}}, \bibinfo
  {author} {\bibfnamefont {M.~E.}\ \bibnamefont {Tai}}, \bibnamefont {et~al.},\
  }\href {\doibase 10.1126/science.1260364} {\bibfield  {journal} {\bibinfo
  {journal} {Science}\ }\textbf {\bibinfo {volume} {347}},\ \bibinfo {pages}
  {1229}} (\bibinfo {year} {2015})\BibitemShut {NoStop}%
\bibitem [{\citenamefont {Kitagawa et~al.}(2012)\citenamefont {Kitagawa},
  \citenamefont {Broome}, \citenamefont {Fedrizzi}, \citenamefont {Rudner},
  \citenamefont {Berg}, \citenamefont {Kassal}, \citenamefont {Aspuru-Guzik},
  \citenamefont {Demler},\ and\ \citenamefont {White}}]{Kitagawa:2012}%
  \BibitemOpen
  \bibfield  {author} {\bibinfo {author} {\bibfnamefont {T.}~\bibnamefont
  {Kitagawa}}, \bibinfo {author} {\bibfnamefont {M.~A.}\ \bibnamefont
  {Broome}}, \bibinfo {author} {\bibfnamefont {A.}~\bibnamefont {Fedrizzi}},
  \bibnamefont {et~al.},\ }\href {http://dx.doi.org/10.1038/ncomms1872}
  {\bibfield  {journal} {\bibinfo  {journal} {Nat. Commun.}\ }\textbf {\bibinfo
  {volume} {3}},\ \bibinfo {pages} {882}} (\bibinfo {year} {2012})\BibitemShut
  {NoStop}%
\bibitem [{\citenamefont {Shindou}(2005)}]{Shindou:2005}%
  \BibitemOpen
  \bibfield  {author} {\bibinfo {author} {\bibfnamefont {R.}~\bibnamefont
  {Shindou}},\ }\href {\doibase 10.1143/JPSJ.74.1214} {\bibfield  {journal}
  {\bibinfo  {journal} {J. Phys. Soc. Jpn.}\ }\textbf {\bibinfo {volume}
  {74}},\ \bibinfo {pages} {1214}} (\bibinfo {year} {2005})\BibitemShut
  {NoStop}%
\bibitem [{\citenamefont {Fu\ and\ Kane}(2006)\citenamefont {Fu}\ and\
  \citenamefont {Kane}}]{Fu:2006}%
  \BibitemOpen
  \bibfield  {author} {\bibinfo {author} {\bibfnamefont {L.}~\bibnamefont
  {Fu}}\ \bibnamefont {and}\ \bibinfo {author} {\bibfnamefont {C.~L.}\
  \bibnamefont {Kane}},\ }\href {\doibase 10.1103/PhysRevB.74.195312}
  {\bibfield  {journal} {\bibinfo  {journal} {Phys. Rev. B}\ }\textbf {\bibinfo
  {volume} {74}},\ \bibinfo {pages} {195312}} (\bibinfo {year}
  {2006})\BibitemShut {NoStop}%
\bibitem [{\citenamefont {Lee et~al.}(2007)\citenamefont {Lee}, \citenamefont
  {Anderlini}, \citenamefont {Brown}, \citenamefont {Sebby-Strabley},
  \citenamefont {Phillips},\ and\ \citenamefont {Porto}}]{Lee:2007}%
  \BibitemOpen
  \bibfield  {author} {\bibinfo {author} {\bibfnamefont {P.~J.}\ \bibnamefont
  {Lee}}, \bibinfo {author} {\bibfnamefont {M.}~\bibnamefont {Anderlini}},
  \bibinfo {author} {\bibfnamefont {B.~L.}\ \bibnamefont {Brown}}, \bibnamefont
  {et~al.},\ }\href {\doibase 10.1103/PhysRevLett.99.020402} {\bibfield
  {journal} {\bibinfo  {journal} {Phys. Rev. Lett.}\ }\textbf {\bibinfo
  {volume} {99}},\ \bibinfo {pages} {020402}} (\bibinfo {year}
  {2007})\BibitemShut {NoStop}%
\bibitem [{\citenamefont {Zhang\ and\ Hu}(2001)\citenamefont {Zhang}\ and\
  \citenamefont {Hu}}]{Zhang:2001}%
  \BibitemOpen
  \bibfield  {author} {\bibinfo {author} {\bibfnamefont {S.-C.}\ \bibnamefont
  {Zhang}}\ \bibnamefont {and}\ \bibinfo {author} {\bibfnamefont
  {J.}~\bibnamefont {Hu}},\ }\href {\doibase 10.1126/science.294.5543.823}
  {\bibfield  {journal} {\bibinfo  {journal} {Science}\ }\textbf {\bibinfo
  {volume} {294}},\ \bibinfo {pages} {823}} (\bibinfo {year}
  {2001})\BibitemShut {NoStop}%
\bibitem [{\citenamefont {Nakajima et~al.}()\citenamefont {Nakajima},
  \citenamefont {Tomita}, \citenamefont {Taie}, \citenamefont {Ichinose},
  \citenamefont {Ozawa}, \citenamefont {Wang}, \citenamefont {Troyer},\ and\
  \citenamefont {Takahashi}}]{Nakajima:2015}%
  \BibitemOpen
  \bibfield  {author} {\bibinfo {author} {\bibfnamefont {S.}~\bibnamefont
  {Nakajima}}, \bibinfo {author} {\bibfnamefont {T.}~\bibnamefont {Tomita}},
  \bibinfo {author} {\bibfnamefont {S.}~\bibnamefont {Taie}}, \bibnamefont
  {et~al.},\ }\href@noop {} {\bibinfo  {journal} {to be published}\
  }\BibitemShut {NoStop}%
\end{thebibliography}

\begin{thebibliography}{4}%
\makeatletter
\providecommand \@ifxundefined [1]{%
 \@ifx{#1\undefined}
}%
\providecommand \@ifnum [1]{%
 \ifnum #1\expandafter \@firstoftwo
 \else \expandafter \@secondoftwo
 \fi
}%
\providecommand \@ifx [1]{%
 \ifx #1\expandafter \@firstoftwo
 \else \expandafter \@secondoftwo
 \fi
}%
\providecommand \natexlab [1]{#1}%
\providecommand \enquote  [1]{``#1''}%
\providecommand \bibnamefont  [1]{#1}%
\providecommand \bibfnamefont [1]{#1}%
\providecommand \citenamefont [1]{#1}%
\providecommand \href@noop [0]{\@secondoftwo}%
\providecommand \href [0]{\begingroup \@sanitize@url \@href}%
\providecommand \@href[1]{\@@startlink{#1}\@@href}%
\providecommand \@@href[1]{\endgroup#1\@@endlink}%
\providecommand \@sanitize@url [0]{\catcode `\\12\catcode `\$12\catcode
  `\&12\catcode `\#12\catcode `\^12\catcode `\_12\catcode `\%12\relax}%
\providecommand \@@startlink[1]{}%
\providecommand \@@endlink[0]{}%
\providecommand \url  [0]{\begingroup\@sanitize@url \@url }%
\providecommand \@url [1]{\endgroup\@href {#1}{\urlprefix }}%
\providecommand \urlprefix  [0]{URL }%
\providecommand \Eprint [0]{\href }%
\providecommand \doibase [0]{http://dx.doi.org/}%
\providecommand \selectlanguage [0]{\@gobble}%
\providecommand \bibinfo  [0]{\@secondoftwo}%
\providecommand \bibfield  [0]{\@secondoftwo}%
\providecommand \translation [1]{[#1]}%
\providecommand \BibitemOpen [0]{}%
\providecommand \bibitemStop [0]{}%
\providecommand \bibitemNoStop [0]{.\EOS\space}%
\providecommand \EOS [0]{\spacefactor3000\relax}%
\providecommand \BibitemShut  [1]{\csname bibitem#1\endcsname}%
\let\auto@bib@innerbib\@empty
%</preamble>
\bibitem [{\citenamefont {Harper}(1955)}]{S:Harper:1955}%
  \BibitemOpen
  \bibfield  {author} {\bibinfo {author} {\bibfnamefont {P.~G.}\ \bibnamefont
  {Harper}},\ }\href {http://stacks.iop.org/0370-1298/68/i=10/a=305} {\bibfield
   {journal} {\bibinfo  {journal} {Proc. Phys. Soc. A}\ }\textbf {\bibinfo
  {volume} {68}},\ \bibinfo {pages} {879}} (\bibinfo {year} {1955})\BibitemShut
  {NoStop}%
\bibitem [{\citenamefont {Roux et~al.}(2008)\citenamefont {Roux}, \citenamefont
  {Barthel}, \citenamefont {McCulloch}, \citenamefont {Kollath}, \citenamefont
  {Schollw\"ock},\ and\ \citenamefont {Giamarchi}}]{S:Roux:2008}%
  \BibitemOpen
  \bibfield  {author} {\bibinfo {author} {\bibfnamefont {G.}~\bibnamefont
  {Roux}}, \bibinfo {author} {\bibfnamefont {T.}~\bibnamefont {Barthel}},
  \bibinfo {author} {\bibfnamefont {I.~P.}\ \bibnamefont {McCulloch}},
  \bibnamefont {et~al.},\ }\href {\doibase 10.1103/PhysRevA.78.023628}
  {\bibfield  {journal} {\bibinfo  {journal} {Phys. Rev. A}\ }\textbf {\bibinfo
  {volume} {78}},\ \bibinfo {pages} {023628}} (\bibinfo {year}
  {2008})\BibitemShut {NoStop}%
\bibitem [{\citenamefont {Kraus\ and\ Zilberberg}(2012)\citenamefont {Kraus}\
  and\ \citenamefont {Zilberberg}}]{S:Kraus:2012a}%
  \BibitemOpen
  \bibfield  {author} {\bibinfo {author} {\bibfnamefont {Y.~E.}\ \bibnamefont
  {Kraus}}\ \bibnamefont {and}\ \bibinfo {author} {\bibfnamefont
  {O.}~\bibnamefont {Zilberberg}},\ }\href {\doibase
  10.1103/PhysRevLett.109.116404} {\bibfield  {journal} {\bibinfo  {journal}
  {Phys. Rev. Lett.}\ }\textbf {\bibinfo {volume} {109}},\ \bibinfo {pages}
  {116404}} (\bibinfo {year} {2012})\BibitemShut {NoStop}%
\bibitem [{\citenamefont {Hatsugai\ and\ Kohmoto}(1990)\citenamefont
  {Hatsugai}\ and\ \citenamefont {Kohmoto}}]{S:Hatsugai:1990}%
  \BibitemOpen
  \bibfield  {author} {\bibinfo {author} {\bibfnamefont {Y.}~\bibnamefont
  {Hatsugai}}\ \bibnamefont {and}\ \bibinfo {author} {\bibfnamefont
  {M.}~\bibnamefont {Kohmoto}},\ }\href {\doibase 10.1103/PhysRevB.42.8282}
  {\bibfield  {journal} {\bibinfo  {journal} {Phys. Rev. B}\ }\textbf {\bibinfo
  {volume} {42}},\ \bibinfo {pages} {8282}} (\bibinfo {year}
  {1990})\BibitemShut {NoStop}%
\end{thebibliography}

\end{document}